\documentclass{article}
\usepackage{graphicx}
\usepackage{amsmath}
\usepackage{amssymb}
\usepackage{longtable}
\usepackage{makecell}
\usepackage{geometry}
\usepackage{threeparttable}
\usepackage{booktabs}

\usepackage[
    backend=biber,
    style=nature,
    sorting=none,
    maxnames=99,
    intitle=true
]{biblatex}
\renewbibmacro*{journal}{%
  \iffieldundef{shortjournal}
    {\printfield{journaltitle}}
    {\printfield{shortjournal}}%
}

\usepackage{hyperref}
\hypersetup{
   colorlinks=true,
   linkcolor=blue,
   filecolor=magenta,      
   urlcolor=cyan,
   citecolor=blue
}

\usepackage{xcolor}

\title{Mesoscale structures in signed networks}

\author{
	Wei Zhang$^{1,2\ast}$,
	Olga Boichak$^{3}$,
    Tristram J. Alexander$^{2,4}$,
    Tiago P. Peixoto$^{5,6,7,8,9}$,\and
	Eduardo G. Altmann$^{1,2}$\and
	\small$^{1}$School of Mathematics and Statistics, University of Sydney, Sydney, 2006 NSW, Australia. \and
    \small$^{2}$Centre for Complex Systems, University of Sydney, Sydney, 2006 NSW, Australia. \and
    \small$^{3}$School of Art, Communication and English, University of Sydney, Sydney, 2006 NSW, Australia.\and
	\small$^{4}$School of Physics, University of Sydney, Sydney, 2006 NSW, Australia. \and
    \small$^{5}$ Center for Critical Computational Studies, Goethe University Frankfurt, 60629 Frankfurt am Main, Germany\and
    \small$^{6}$ Institute of Computer Science, Goethe University Frankfurt, 60629 Frankfurt am Main, Germany\and
    \small$^{7}$IT:U Interdisciplinary Transformation University, 4040 Linz, Austria.\and
	\small$^{8}$ Department of Network and Data Science, Central European University, 1100 Vienna, Austria\and
    \small$^{9}$ Complexity Science Hub, 1030 Vienna, Austria\and
    \small$^\ast$Corresponding author. Email: wei.zhang4@sydney.edu.au\and
}

\date{}

\begin{document}

\maketitle

\begin{abstract}

Mesoscale structures in signed networks have been studied under the limiting assumption of the validity of social balance theory, which predicts positive connections within groups and negative connections between groups.
Here, we propose and apply a methodology that overcomes this limitation and is able to find and characterize also the different possible unbalanced structures in signed networks. 
Applying our methodology to 24 empirical networks, from social-political, financial, and biological domains, we find that unbalanced mesoscale structures are prevalent in real-world 
networks, including 
cases with substantial balance at the microscale of triangles.
In particular, we find that assortativity often prevails regardless of the interaction sign and that core-periphery structures are typical in online social networks.
Our findings highlight the complexity of mesoscale relational structures, the importance of using computational methods that are a priori agnostic to specific patterns, and the importance of independently evaluating micro- and mesoscale predictions of social balance theory.

\end{abstract}

\textbf{Teaser:} Looking beyond balance: mesoscale patterns in signed networks show richer structure.

\section{Introduction}

The recent renewed interest in signed networks~\cite{ordozgoiti2020finding,ruiz2023triadic,pougue2023learning,talaga2023polarization,hao2024proper,gallo2024assessing,gallo2025patterns,fernando2025signed,gonzalez2025defines} reflects the need for network representations of complex phenomena to distinguish between opposing types of interactions. For instance, online interactions among users (vertices) do not always reflect friendship or agreement (positive edges) and are thus better described through networks that account for enmity or disagreement (negative edges). Representing these relationships as signed networks allows researchers not only to capture the presence and strength of connections among vertices, but also differentiate the effects of positive and negative interactions on the structure of the network. Large signed networks typically show a rich combination not only of microscale (e.g., triadic interactions) but also of mesoscale structures (e.g., communities), that are known to be essential to understand the underlying complex system (e.g., the partition of users in opposing groups)~\cite{fortunato2010community,traag2009community,mucha2010community,peixoto2015inferring}.

Signed networks have been primarily studied within \textit{structural balance theory} \cite{cartwright1956structural}, which originated in studies of social and economic networks \cite{heider1946attitudes}. A signed network is considered balanced (anti-balanced) if and only if all cycles have an even number of negative (positive) edges \cite{cartwright1956structural,harary1957structural,tian2024spreading}. In the simplest case of 3-cycles, the signed network with all triads having zero or two negative edges is balanced. While this microscale perspective of structural balance theory is the most popular, structural balance can also be equivalently viewed from a mesoscale perspective: balanced (anti-balanced) signed networks can be partitioned into 2 communities whose intra-links are all positive (negative) and inter-links are all negative (positive) \cite{harary1953notion,harary1957structural}. In a weaker version, k-balance theory states that networks are balanced if they can be partitioned into $k \geq 2$ communities with all intra-community links being positive and all inter-community links being negative~\cite{doreian1996partitioning,gallo2025patterns,fernando2025signed}.
These balance conditions are rarely satisfied strictly in signed networks obtained from empirical datasets. Approaches to quantify the extent into which a signed network is balanced involve the comparison to suitable null models~\cite{Facchetti2011,hao2024proper} of balance measures computed at different network scales~\cite{aref2020multilevel,talaga2023polarization}.
At a micro scale, the {\it degree of balance} (DoB) is computed based on local configurations, including triadic or motif analysis~\cite{cartwright1956structural,singh2017measuring}; at a meso scale, the focus shifts to the aggregate structure of the network and its distance from a perfectly balanced state, quantified by the {\it frustration index} (FI)~\cite{Facchetti2011,aref2019balance,gallo2024assessing};
structural balance can also be viewed from a dynamical perspective, and quantified using the spectrum of the signed Laplacian~\cite{kunegis2010spectral,cucuringu2019sponge} and walk-based measure~\cite{estrada2014walk,kirkley2019balance}.

A critical step in the mesoscale approach mentioned above is the partitioning of the network into groups. Different community detection methods, originally proposed for simple networks, have been recently extended to signed networks: modularity maximization~\cite{newman2004finding,traag2009community,mucha2010community,amelio2013community,traag2015detecting,pougue2024signedlouvain}, spectral clustering \cite{kunegis2010spectral,chiang2012scalable,kunegis2014applications,cucuringu2019sponge}, and spin-glass \cite{reichardt2006statistical,facchetti2012exploring}. 
The primary goal of these methods is to find a partition that is close to a k-balanced partition, i.e., to divide the network into separate communities so that nodes within the same community have as many positive connections as possible, while nodes in different communities have as many negative connections as possible.
These descriptive partitioning methods are designed to optimize for specific structural objectives~\cite{traag2009community,aref2020multilevel,talaga2023polarization} and can succeed when structural balance is the dominant feature shaping the mesoscale organization of the network. However, a substantial challenge arises when the goal is the unbiased detection of mesoscale structures without the {\it a priori} assumption of structural balance.
Because these methods are constrained by their objective functions, they may impose balanced partitions even when the data is better described by alternative patterns or, as we show below, even in random networks where balanced structures are absent.
It is thus unclear in which extent the reported balanced structures are the dominant mesoscale feature of the network or simply a consequence of the assumptions made by the methods.
Since mesoscale structures in signed networks have been studied almost exclusively within structural balance theory, several fundamental questions remain open:
To what extent are the mesoscale structures of real-world signed networks aligned with structural balance theory?
Are there different types of unbalanced structures, and if so, what do they reveal about the organization of large scale signed networks?

In this work, we propose and apply a framework to overcome the limitations of previous approaches and to answer these questions. Our goal is to find, characterize, and quantify the strength of mesoscale structures in signed networks beyond the balance assumption. 
We achieve this by proposing an exhaustive pairwise classification of structures in signed graphs -- generalizing previous classifications for unsigned graphs~\cite{betzel2018,liu2023nonassortative} and connecting this classification to structural balance theory -- and combining it with state-of-the-art inferential methods for community detection~\cite{peixoto2023descriptive}, which have already highlighted that diverse mixing patterns may simultaneously coexist in a network~\cite{jiang2015stochastic,Leto2017,guimera2020one,peel2022statistical,peixoto2023implicit,gallo2024assessing}.
In particular, we use stochastic block models (SBMs) \cite{karrer2011stochastic,peixoto2015inferring,peixoto2018nonparametric,peixoto2019bayesian,peixoto2023descriptive} that leverage statistical evidence and do not a priori adhere to specific types of balanced community structures.
Applying this framework to 24 empirical datasets, we find that unbalanced structures are present in almost all networks and are dominant in a significant portion of them. In particular, we find that social networks are often best partitioned into groups in which positive and negative links happen within the same group, possibly reflecting the interaction opportunities available to individuals. We also find that relationships akin to core-periphery structures are widespread in (the negative edges of) online social networks. 
Finally, we show how the micro and mesoscale characterization of structural-balance theory often leads to complementary pictures of the processes shaping complex signed networks.

\section{Results}

\subsection{Methodology}\label{sec:Methodology}

\begin{figure}[ht!]
\centering
\includegraphics[width=0.9\linewidth]{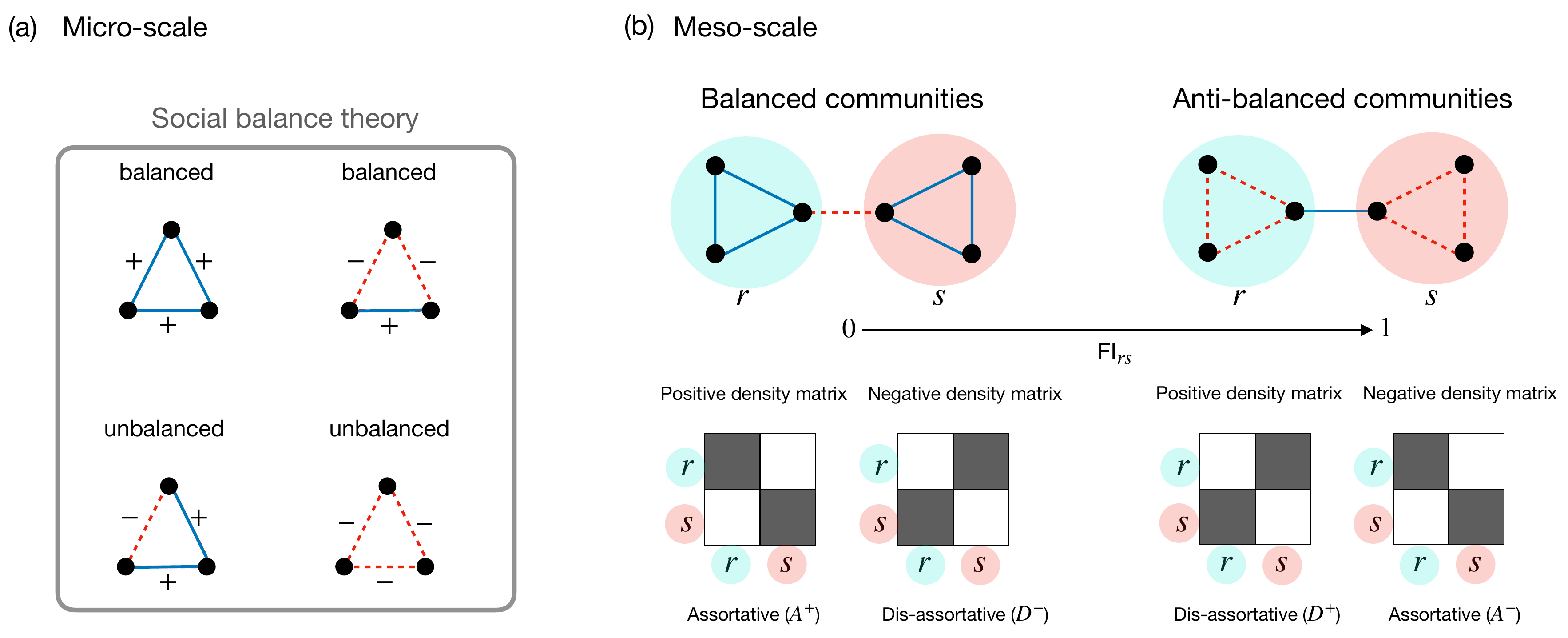}
\caption{\textbf{Structural balance theory from (a) microscale and (b) mesoscale perspectives.}
(a) shows fundamental triads or motifs considered as balanced and unbalanced by social balance theory.
(b) Balanced (left) and anti-balanced (right) communities. The top panel shows simple signed networks, with positive links as solid lines and negative links as dashed lines. The two groups $r$ and $s$ are marked with circles. The bottom panels show the corresponding edge density matrices~\eqref{eq.w}. 
}
\label{fig:rela}
\end{figure}

\begin{figure}[ht!]
\centering
\includegraphics[width=0.9\linewidth]{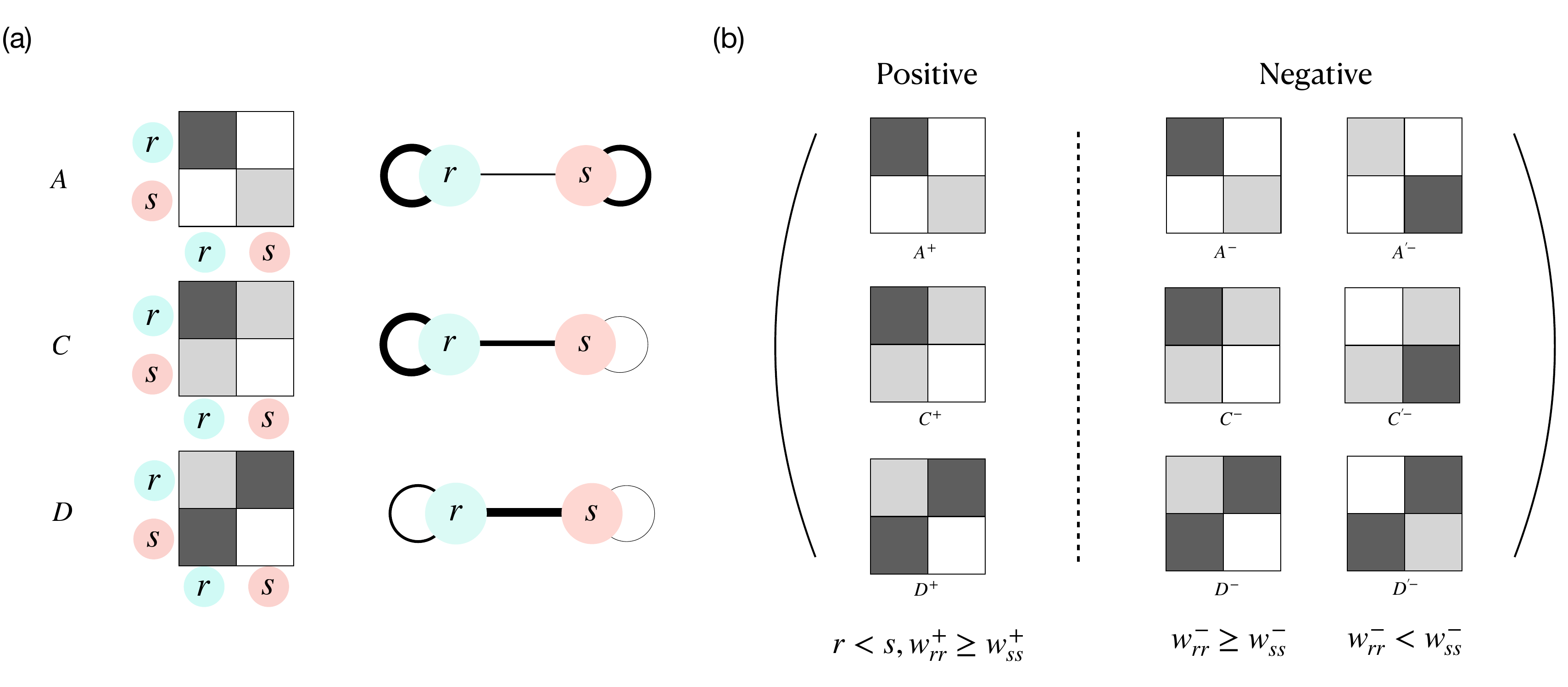}
\caption{\textbf{Pairwise classification of community types}. (a) The three potential pairwise-community relations between communities $r$ and $s$ are represented by density matrices. The darker blocks indicate a higher density of edges.
Assortative $A^{\pm}$ (disassortative $D^{\pm}$) cases have the two largest entries in the diagonal $w_{rr}^{\pm}$, $w_{ss}^{\pm}$ (off-diagonal $w_{rs}^{\pm}$,$w_{sr}^{\pm}$). The core–periphery type of relationship $C^{\pm}$ has one sparser community (periphery) linked strongly to the denser community (core), which is in turn more interconnected within itself.
(b) The three possible types for positive links and six for negative links, thus leading to a total of 18 possible combination types. Note, blocks are arranged such that the left column has the highest density of positive intra-block edges. This freedom is unavailable for the negative edges, so there are twice as many types.}
\label{fig:class1}
\end{figure}

\begin{figure}[ht!]
\centering
\includegraphics[width=0.85\linewidth]{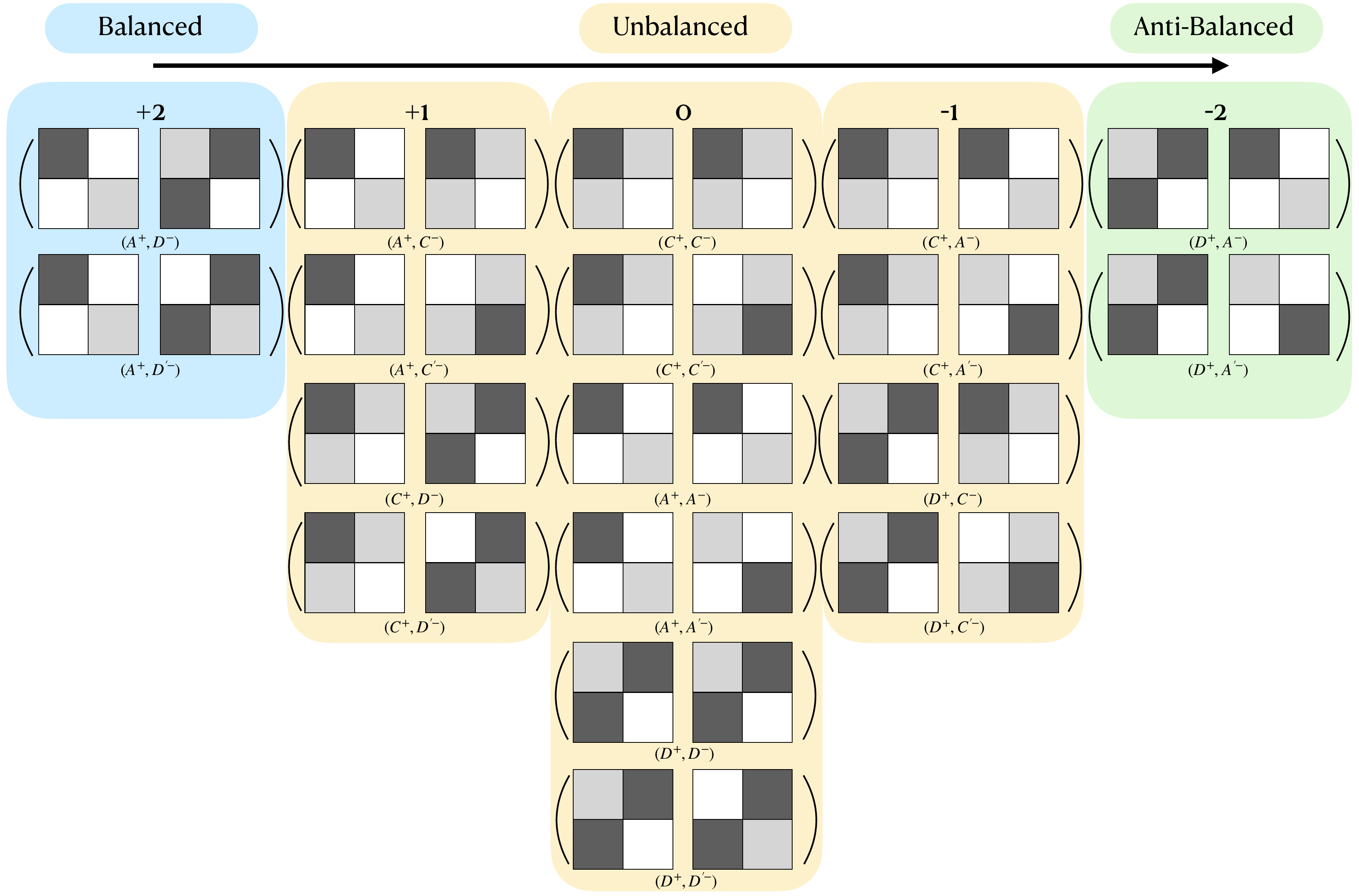}
\caption{\textbf{Pairwise classification of mesoscale structures in signed networks.}
`$A^{\pm}$', `$C^{\pm}$' and `$D^{\pm}$' stand for `assortative', `core–periphery' and `disassortative' in positive ($+$) or negative ($-$) interaction, respectively, and are defined by Eq.~\eqref{eq.adc}. 
For example, the pair $(A^+, D^{'-})$ represents the positive edges are assortative and the negative edges are disassortative, and the mark $'$ labeled in $D^{'-}$ indicates that the negative diagonal entries $w_{rr}^- < w_{ss}^-$. From left to right, the 18 different possible configurations are divided according to their level of balance. A relationship is balanced when positive edges are assortative (dense within communities) and negative edges are disassortative (dense between communities).
In contrast, a structure is anti-balanced when positive edges are disassortative and negative edges are assortative. These balanced (anti-balanced) structures are considered to have weak structural balance (anti-balance) since they allow the limited presence of other entries in the density matrix. The remaining unbalanced configurations are scored as follows: assortative is $+1$, core-periphery is $0$, and disassortative is $-1$. For negative edges, the scoring is: disassortative is $+1$, core-periphery is 0, and assortative is $-1$. 
}
\label{fig:class2}
\end{figure}

We first consider how mesoscale partitions of signed networks are related to structural balance theory at its usual microscale representation. Figure~\ref{fig:rela} shows a simple case connecting the two scales, with balanced and anti-balanced mesoscale communities fulfilling the microscale definitions based on triads.
A balanced network shows diagonal blocks in the positive edge density matrix and off-diagonal blocks in the negative edge density matrix (bottom of Fig.~\ref{fig:rela}b).
This indicates that balanced communities exhibit assortative relationships in positive interactions and disassortative relationships in negative interactions~\cite{betzel2018,liu2023nonassortative}. Conversely, anti-balanced communities show disassortative relationships in positive interactions and assortative relationships in negative interactions. This motivates us to use the edge density matrix to measure and classify diverse community structures, including unbalanced ones.

Consider a simple undirected signed graph $G$ with adjacency matrix $(A_{ij}) \in \mathbb{R}^{N \times N}$, where each entry $A_{ij}$ can be written as the sum of two disjoint components $A_{ij} = A_{ij}^+ + A_{ij}^-$ with $A_{ij}^+ >0$ represents positive edges and $A_{ij}^- <0$ represents negative edges. The graph is partitioned into $r = 1,2,3,...,B$ disjoint subgroups, each containing $N_r$ nodes. 
For a pair of communities $r$ and $s$, the density of positive interactions and negative interactions is given by the ratio of existing edges and possible edges as

\begin{equation}\label{eq.w}
w_{rs}^{\pm} = 
\begin{cases}
\sum_{i\in r, j\in s}\frac{|A_{ij}^{\pm}|}{N_rN_s}, & \text{if } r \neq s,\\
\sum_{i\in r, j\in s}\frac{|A_{ij}^{\pm}|}{N_r(N_s-1)}, & \text{if } r = s.
\end{cases}
\end{equation}

We now classify the relationship between a pair of communities $r$ and $s$ based on $w_{rs}^{\pm}$. In simple graphs, as shown in Fig.~\ref{fig:class1}a, this relationship can be classified as either assortative, core-periphery, or disassortative~\cite{rombach2017core,betzel2018,liu2023nonassortative}. 
In signed networks, these relationships apply to both positive and negative edges:
\begin{equation}\label{eq.adc}
\begin{cases}
\text{Assortative} (A^{\pm}), & \text{if } w_{rs}^{\pm} < \text{min}(w_{rr}^{\pm},w_{ss}^{\pm}), \\
\text{Core-periphery} (C^{\pm}), & \text{if } \text{min}(w_{rr}^{\pm},w_{ss}^{\pm})  < w_{rs}^{\pm}< \text{max}(w_{rr}^{\pm},w_{ss}^{\pm}), \\
\text{Disassortative} (D^{\pm}), & \text{if } w_{rs}^{\pm} > \text{max}(w_{rr}^{\pm},w_{ss}^{\pm}).
\end{cases}
\end{equation}
In the classification above, we adopt the community label convention that $w_{rr}^{\pm} \geq w_{ss}^{\pm} $ for $ r < s $. By applying this convention to $w^+$, this reduces the number of ranked relationships for the positive edges to three (shown in Fig.~\ref{fig:class1}a), while six options remain for negative matrix $w^-$ (shown in Fig.~\ref{fig:class1}b), because the rankings of $ w_{rr}^-$ and $ w_{ss}^-$ may be oppositely ordered. To differentiate the two options (i.e., $w_{rr}^- > w_{ss}^-$ and $w_{rr}^- < w_{ss}^-$), we label this latter configuration by a prime symbol $'$.
For instance, the pair $ (C^+, C^{'-}) $ signifies that the two communities are in a core-periphery relationship in both positive and negative edges, but that the community that acts as core is {\it different} in each of the edge types
(The notation $(C^+,C^-)$ represents the relationship in which both cores are the same community). When no distinction between these cases is intended, we denote the relationship as $(C^+,C^{(')-})$.
To quantify mesoscale balance, we define a structural scale connecting the two theoretical extremes: Balanced ($A^+,~D^{(')-}$), and Anti-balanced ($D^+,~A^{(')-}$). Each pairwise relationship type is assigned a score: $+1$ for assortative ($A$), $-1$ for disassortative ($D$), and $0$ for core-periphery ($C$) in the case of positive interactions (these values are reversed for negative interactions).
Thus, there are a total of $ 3 \times 6 =18$ possible relationships between two communities $r$ and $s$ for signed networks. These relationships can be grouped into balanced, unbalanced, and anti-balanced groups, as shown in Fig.~\ref{fig:class2}.

\begin{figure}[ht!]
\centering
\includegraphics[width=0.75\linewidth]{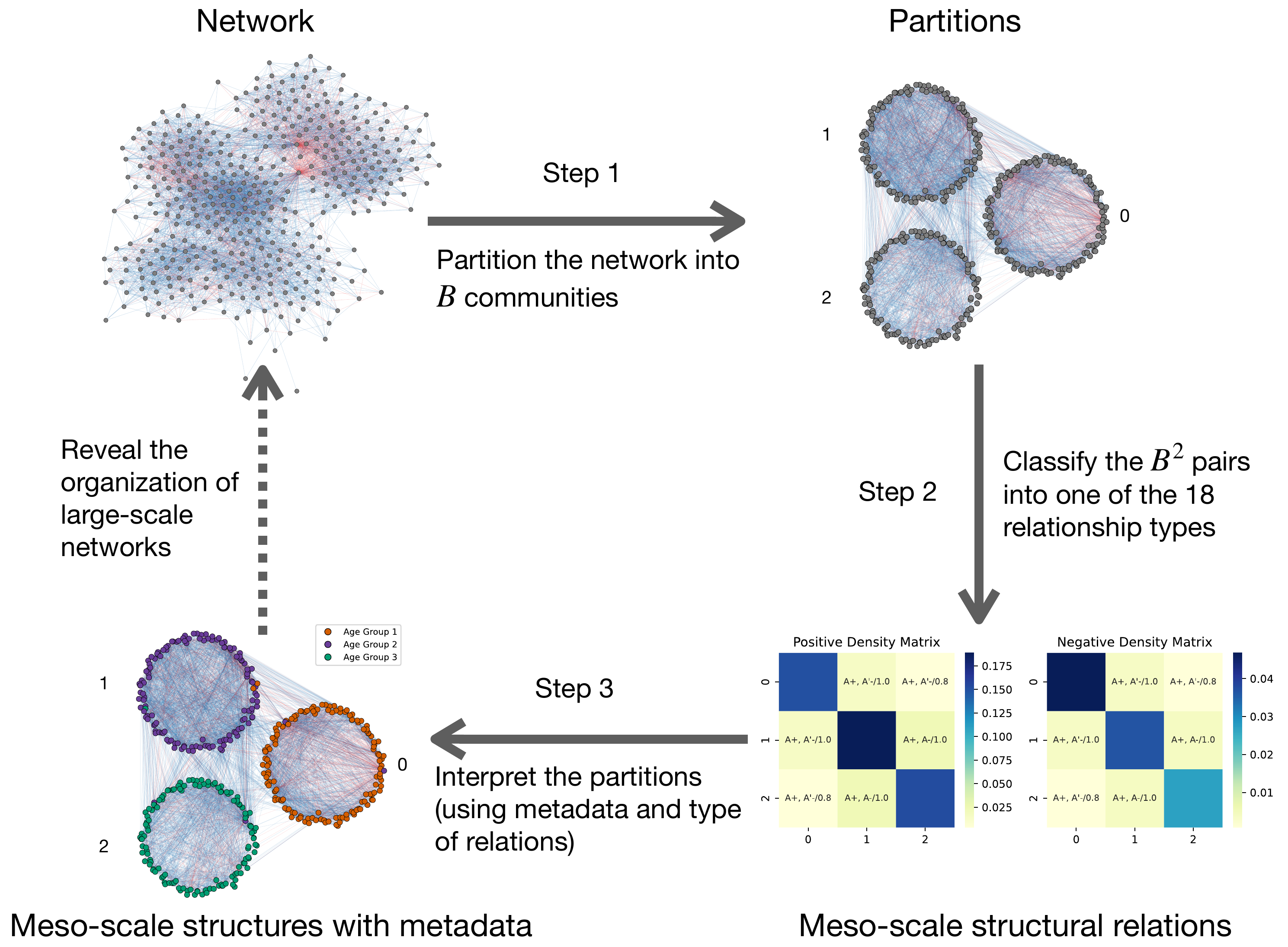}
\caption{\textbf{Illustration of the main steps of our methodology.}
Step 1: The signed network is partitioned into $B$ communities using a suitable partition method, such as the ones listed in Table~\ref{tab:methods}.
Step 2: Each pair of communities is classified according to its type, as proposed in Fig.~\ref{fig:class2}. 
Step 3: This classification is then used -- in combination with metadata, when available -- to interpret the mesoscale structures and quantify the degree of balance at a mesoscale level.
}
\label{fig:workflow}
\end{figure}

\begin{table}[ht!]
\centering
\begin{threeparttable}
\caption{\textbf{List of methods used for the partition of signed networks.}
The table compares five methods according to their core idea, whether they assume structural balance, their computational complexity, and their empirical runtime on three example networks of various sizes: (i) $N=446$ with $13,908$ positive and $36,780$ negative edges; (ii) $N=3,883$ with $2,808$ positive and $2,672$ negative edges; (iii) $N=8,711$ with $22,910$ positive and $2,163$ negative edges) using a laptop (with an Apple M3 chip, 16 GB of memory, running macOS 26.3 25D125). For smaller numbers of groups $B$, methods with higher asymptotic complexity (e.g., spectral clustering) can run faster than inferential alternatives, see Ref.~\cite{peixoto2023descriptive}}.
\label{tab:methods}
\begin{tabular}{
>{\raggedright\arraybackslash}p{0.15\textwidth}
>{\centering\arraybackslash}p{0.14\textwidth}
>{\centering\arraybackslash}p{0.14\textwidth}
>{\centering\arraybackslash}p{0.14\textwidth}
>{\centering\arraybackslash}p{0.14\textwidth}
>{\centering\arraybackslash}p{0.14\textwidth}
}
\toprule
\textbf{Property} 
& \textbf{LSBM}~\cite{peixoto2015inferring} 
& \textbf{WSBM}~\cite{peixoto2018nonparametric} 
& \textbf{Louvain}~\cite{traag2015detecting,pougue2024signedlouvain} 
& \textbf{Spectral}~\cite{cucuringu2019sponge} 
& \textbf{Spin-glass}~\cite{traag2009community} \\
\midrule
\textbf{Core Idea} 
& Inference: Two-layer shared partition. 
& Inference: Block-pair distributions. 
& Modularity: Iterative modularity maximization. 
& Linear algebra: Maximizes internal (+), external (-) edges.
& Statistical-physics: Spin alignment via edge signs. \\
\midrule
\textbf{Balance Assumption?} 
& No & No & Yes & Yes & Yes \\
\midrule
\textbf{Complexity} 
& $O(N\log^2 N)$ 
& $O(N\log^2 N)$ 
& $O(N\log N)$ 
& $O(N^2\log N)$ 
& -- \\
\midrule
\multicolumn{6}{l}{\textbf{Running time in seconds (number of groups $B$):}} \\
\midrule
\textbf{Network i} 
& 3.64s (19) & 2.78s (27) & 0.10s (3) & 0.23s (2) & 40.25s (12) \\
\textbf{Network ii} 
& 2.88s (4) & 2.26s (8) & 0.06s (1911) & 3.47s (2) & 27.00s (50) \\
\textbf{Network iii} 
& 8.39s (35) & 8.96s (36) & 0.17s (325) & 24.36s (13) & 27.00s (49) \\
\bottomrule
\end{tabular}
\end{threeparttable}
\end{table}

Our classification serves as building blocks to analyze the structural profile of a signed network.
In order to apply the pairwise classification described above to a network with $B\geq2$ communities, we need to (i) identify the groups of nodes (i.e., partition the network into $B$ communities); and (ii) classify each of the $\binom{B}{2}$ pairs' relation type into one of the 18 possible combinations, as shown in Fig.~\ref{fig:workflow}.
Different methods for network partition can produce varying $ B \times B $ density matrices, and our methodology can be applied to any of these partition methods. 
The results presented here primarily involve five methods,
including traditional methods that focus on balanced partitions -- Louvain~\cite{traag2015detecting,pougue2024signedlouvain}, spectral clustering~\cite{cucuringu2019sponge}, and spin-glass~\cite{traag2009community}-- and inferential approaches that are agnostic about it -- Layered SBM (LSBM)~\cite{peixoto2015inferring} and Weighted SBM (WSBM)~\cite{peixoto2018nonparametric} (comparison of the methods see Table~\ref{tab:methods}). 
Hierarchical versions of SBMs~\cite{peixoto2014hierarchical} can also be directly used in combination with our classification by analyzing pairwise relationships at different hierarchy levels.
We then analyze each of the $\binom{B}{2}$ pairs, we classify their relationship type into one of the 18 possible combinations, and we focus on whether a specific relationship appears in that network (contributing to the overall prevalence of the type) and which type is the most common (dominant).

\begin{figure}[ht]
\centering
\includegraphics[width=1\linewidth]{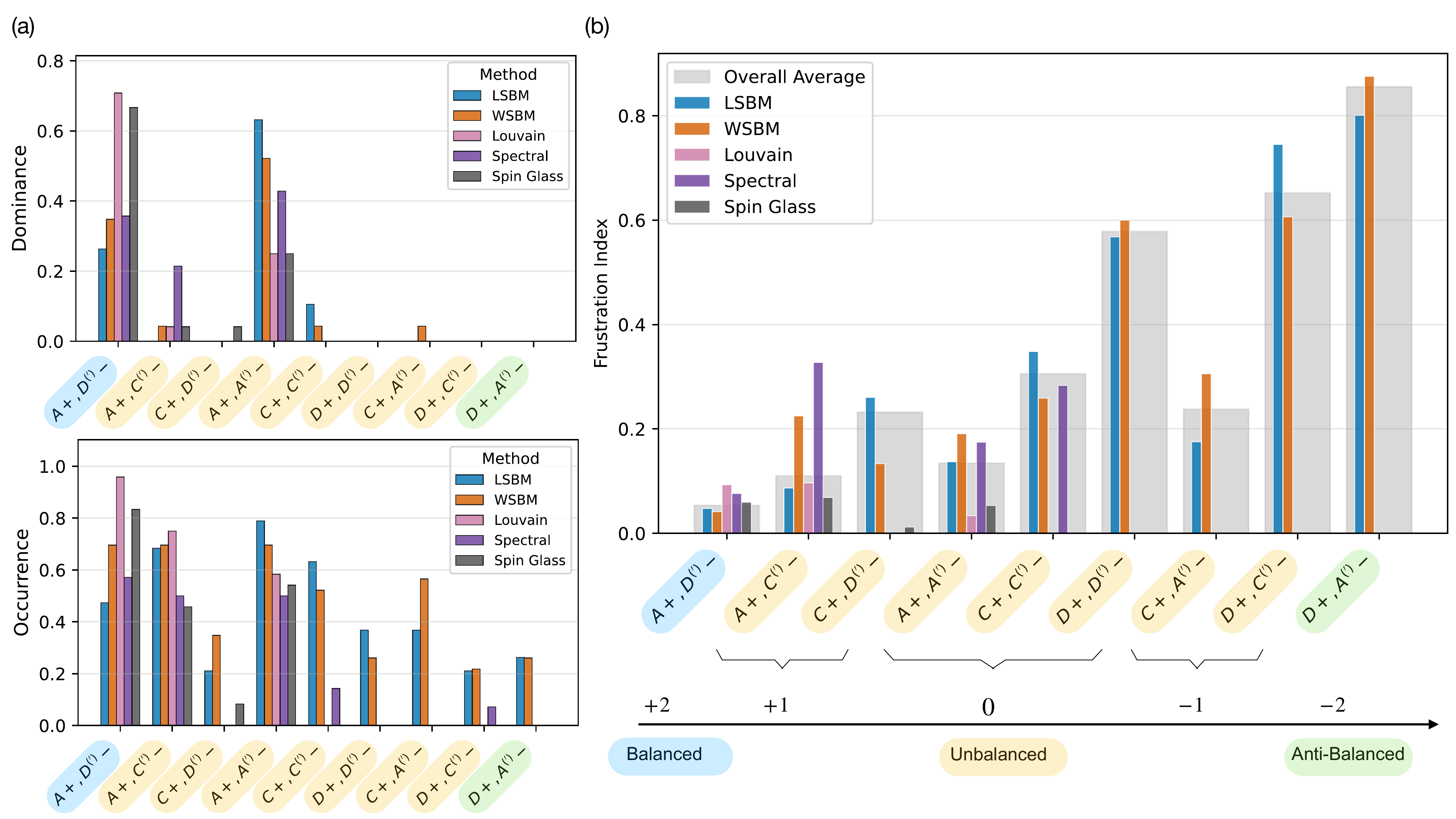}
\caption{\textbf{Survey on 24 empirical networks.} (a) Community types in empirical networks. Top: fraction of networks in which each community type is dominant (computed as the proportion of networks in which the type is the most prevalent). Bottom: fraction of networks in which each community type occurs (computed as the proportion of networks in which the type appears at least once). (b) Frustration index in empirical networks. The x-axis represents balance categories ranging from balanced, unbalanced, to anti-balanced configurations, while the y-axis shows the frustration index. Bars correspond to different methods, with the overall average shown in gray. The calculation of the pairwise frustration index refers to Methods~\ref{FI}.}
\label{fig:1}
\end{figure}

\subsection{Unbalanced structures are common}
\label{sec:1}

We now apply the methodology described in Sec.~\ref{sec:Methodology} to empirical networks. We collected 24 empirical signed networks, which include 18 social-political networks, 2 financial networks, and 4 gene-regulatory networks (see Methods~\ref{data} and Supplementary Materials for a detailed description of the data set). These networks vary substantially in size, between 16 and 10,884 nodes and between 40 and 251,406 edges (see Table~\ref{tab:descriptive} for a list).

The results obtained across all networks are summarized in Fig.~\ref{fig:1}. 
As expected, the dominant relationship type in most networks for the balanced-based methods (e.g., Louvain and spin-glass) is the balanced structure $(A^+, D^{(')-})$, with few unbalanced structures detected except for $(A^+, A^{(')-})$. This reflects the fact that these methods aim to find balanced partitions, thus providing little information about the type of structure present in the data. More interestingly, the results obtained for the two inferential methods (i.e., LSBM and WSBM), which do not focus exclusively on balanced partitions and rely instead on statistical evidence, find more unbalanced structures with the dominant type of relationship over all networks being the unbalanced structure $(A^+, A^{(')-})$ and with different core-periphery relationships observed as well.
Similar results are observed when we restrict the analysis to social-political networks, financial networks, and gene-regulatory networks (Figs.~\ref{fig:S1}-\ref{fig:S3}).
Notably, non-social mechanisms shape structure differently from social networks and are known to show lower degrees of balance~\cite{aref2019balance}. Indeed, for gene-regulatory networks, our method shows (Fig.~\ref{fig:S2}) that both SBM methods identify unbalanced structural relationships to be dominant in all cases, contrary to the Louvain method that detects balanced structures as dominant in all cases.
This finding indicates that unbalanced structures among communities are not just a theoretical possibility; they are actually common in empirical networks, provided suitable partition methods are used. 

We validate our pairwise approach by comparing the consistency of the relationships we found with traditional methods from structural balance theory. We compare our classification of mesoscale structure types to the average pairwise frustration index, which measures the fraction of misplaced links, positive interlinks and negative intralinks, between communities (Methods~\ref{FI} for details). A frustration index $\text{FI}=0$ indicates perfect balance, while $\text{FI}=1$ indicates complete anti-balance. The results in Fig.~\ref{fig:1}b reveal a clear trend: balanced structures correspond to low frustration, unbalanced structures exhibit intermediate levels, and anti-balanced structures show the highest frustration. This progression is consistent across different methods, and the overall averages align closely with theoretical expectations, confirming the alignment of our classification scheme to structural balance theory.

Below we explore the significance and interpretation of these unbalanced relationships, focusing on the partitions obtained using SBMs. This choice is motivated by their ability to detect such patterns, as the underlying generative network model can, in principle, contain any of the pairwise relationship types discussed previously. We use degree-corrected SBMs as a generative process to prevent the classification of high-degree nodes into blocks based solely on their connectivity~\cite{karrer2011stochastic}. The number of blocks (communities) $B$ and the block allocation of each node are obtained by performing Bayesian inference, following the approach proposed in Refs.~\cite{peixoto_graph-tool_2014,peixoto2019bayesian,peixoto_merge_2020}. Sampling the posterior distribution over partitions allows us to evaluate the robustness of the proposed partitions and of the classification of the mesoscale structures, including the extent into which they are balanced (see Methods~\ref{ssec:sbm} for details). 
The results of posterior samplings (Supplementary Materials, SM-Tables~\ref{tab:LSBM}-\ref{tab:WSBM} and Fig.~\ref{fig:S5}) confirm that unbalanced structures were predominant.
We compared the dominant types identified in our survey with those from posterior sampling. For networks with dominant-type certainty $> 0.5$, the consistency rate was 100\% for LSBM and 88\% for WSBM; across all networks, it was 96\% and 79\%, respectively.
Inconsistencies arise from two main factors: low posterior certainty, indicating alternative community structures, and cases where the survey-identified type is the second most frequent, with a frequency close to that of the dominant type (SM Fig.~\ref{fig:S6}).

\subsection{Assortativity prevails regardless of interaction sign}

\begin{figure}[ht]
\centering
\includegraphics[width=0.9\linewidth]{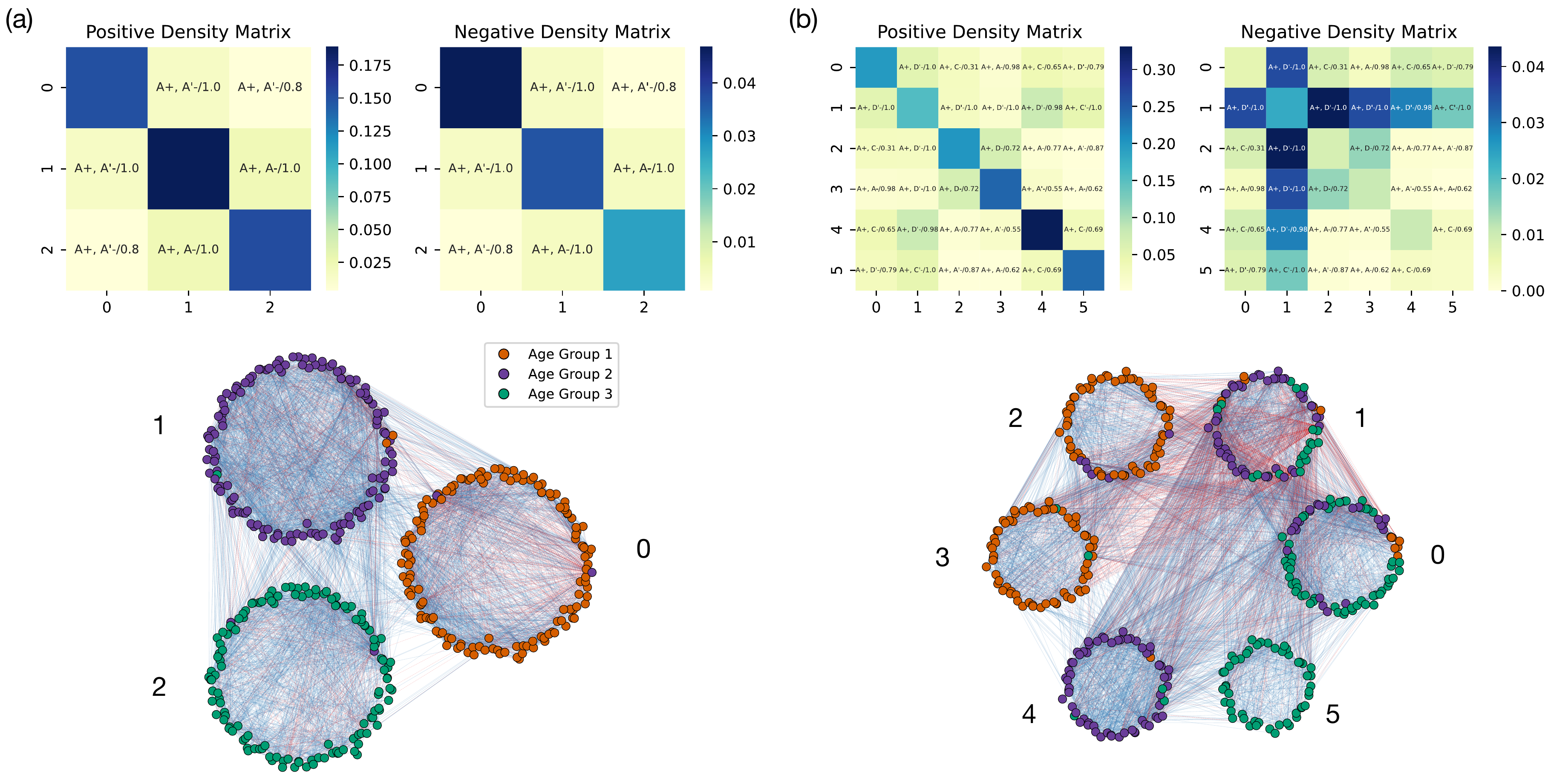}
\caption{\textbf{Partition of Spanish High School network with (a) WSBM and (b) Louvain.} Top: The blocks are ranked according to community size. The labels in each entry of the matrix correspond to the structure type and the robustness score (see Methods~\ref{Robustness}). Bottom: The communities are grouped closely together, and the numbers denote the labels of each group. The colors of the nodes represent different age group memberships.}
\label{fig:spanish}
\end{figure}

\begin{figure}[h!]
\centering
\includegraphics[width=1\linewidth]{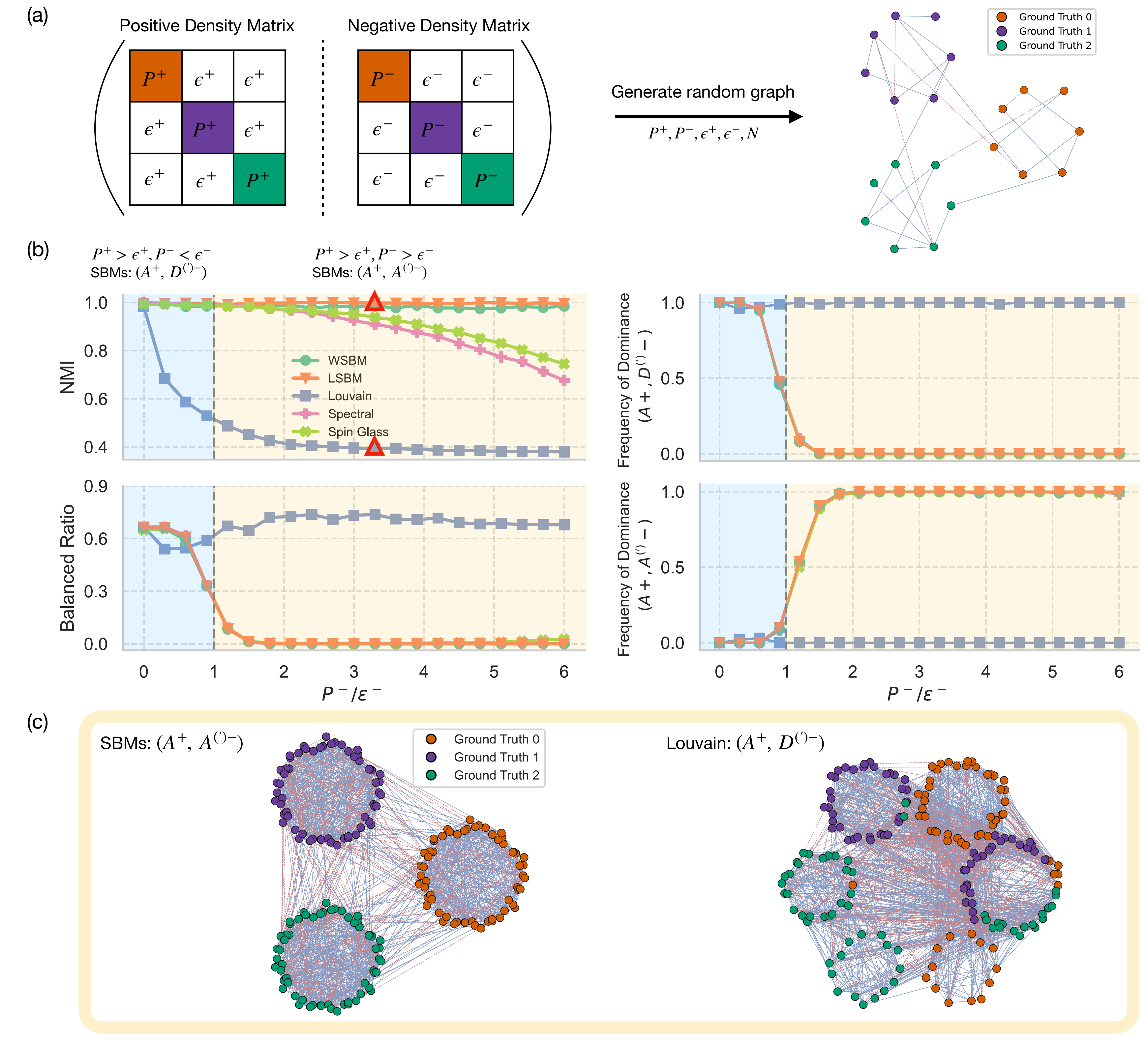}
\caption{\textbf{Heuristic methods lead to biased partitions in unbalanced synthetic networks.} 
(a) Generative model schematic with three ground-truth groups with equal size ($N/3$ nodes each). The intra-group interaction probability is $P^+$ (positive) and $P^-$ (negative), and the inter-group probabilities are $\epsilon^+$ (positive) and $\epsilon^-$ (negative). Nodes are color-coded by group, with solid blue lines and red dashed lines representing positive and negative links, respectively. For $P^- < \epsilon^-$, the network is dominated by $(A^+, D^{(')-})$ relationships (balanced regime) while for $P^- > \epsilon^-$ by $(A^+, A^{(')-})$ relationships. 
(b) Performance of different network-partition methods against the ground truth. Normalized Mutual Information (NMI, Methods~\ref{NMI}) and Balanced Ratio $(A^+, D^{(')-})$ (left panels), and fraction of random networks identifying specific dominant interaction types (right panels), across 100 independent synthetic random networks. The dashed grey line ($P^- = \epsilon^-$) marks the theoretical boundary between the balanced (blue) and unbalanced (yellow) regimes. SBMs correctly identify the transition from balanced to unbalanced dominance, matching the generative ground truth. In contrast, Louvain always finds balanced structures, even when the ground truth is unbalanced. 
(c) Example of partitions obtained by SBM and Louvain, with nodes colored according to the ground truth. The parameters used in all plots are $P^+ = 0.2, \epsilon^+ = \epsilon^- = 0.01, N = 180$. In panel (c), $P^-=0.033$ was used (red triangle in panel b).}
\label{fig:syn}
\end{figure}

In most empirical networks investigated in Sec.~\ref{sec:1}, the dominant type found using SBMs is the unbalanced structure $(A^+, A^{(')-})$, while for other methods, the dominant type is the balanced structure $(A^+, D^{(')-})$.
To understand this difference and the origin of $(A^+, A^{(')-})$, we now focus on the example of the relationship between students across the three years of high-school in a Spanish school.
The results in Fig.~\ref{fig:spanish} show that, without using any age information, the partitions obtained from SBMs align with the three distinct age groups with resulting assortative relationships between any two groups, in both positive and negative interactions $(A^+, A^{(')-})$.
The robustness of this interaction type and our conclusions is further confirmed by our posterior sampling analysis (see SM-Table~\ref{tab:spanish}).
This illustrates a pattern where links are driven by the interaction of students within their age groups, regardless of the signs of the ties, while interactions between different age groups are relatively scarce.
While social homophily is recognized as the foundational mechanism of connectivity in positive-only networks~\cite{mark2010}, our findings reveal that negative interactions adhere to the same structural logic.
In comparison, the groups detected by Louvain involve a mixture of age groups and are mostly in balanced relationships ($A^+, D^{(')-}$) with each other.

To validate this finding, we generate synthetic random networks mimicking high-school interactions. We consider three equal-sized age groups with positive edges predominantly within the group and negative edges controlled by a tunable parameter. Our results in Fig.~\ref{fig:syn} show that, when the negative edges are across different groups, all methods correctly identify the (ground-truth) balanced partition. Increasing the number of negative edges within groups, the SBM methods succeed in retrieving the three groups, but the heuristic methods have a much more limited success. Most strikingly, the Louvain method finds a larger number of groups that show balanced relationships, in line with the observations in the Spanish high-school data. 
Altogether, these results show that assortativity (regardless of interaction sign) is a prevailing pattern in signed networks. This pattern is masked by the use of heuristic descriptive methods (such as Louvain), which find balanced network partitions $(A^+, D^{(')-})$ that can be wrongly interpreted as a confirmation of the predictions of social balance theory.

A possible origin of $(A^+,A^{(')-})$ structures, such as the one reported in the Spanish High School and synthetic network, is a network with assortative communities with edge signs added independently of the mesoscale structure. In this trivial case, the mesoscale partition would be retrieved using classical community-detection methods (e.g., classical Louvain). To test whether such a trivial explanation applies in our case, we used description length to compare our WSBM model of sign graphs with a simpler SBM model for the unsigned graph (see Methods Sec.~\ref{model selection} for details). The results obtained in the Spanish High School indicate that the signed-graph WSBM is overwhelmingly preferred, indicating that despite the prevailing $(A^+,A^{(')-})$ structure, there is still significant alignment between node connectivity and edge signs at the mesoscale level. Applying this comparison to the $24$ networks used in our survey, we find that $18$ of them ($75\%$) show this preference, while in $6$ ($25\%$) the simpler model of unsigned SBM is preferred (see SM-Fig.~\ref{fig:S7}). This shows that edge signs are highly informative about the mesoscale organization in most empirical networks, including cases with both balanced and unbalanced partitions.

\subsection{Core-periphery is typical in social networks}

\begin{figure}[ht]
\centering
\includegraphics[width=1\linewidth]{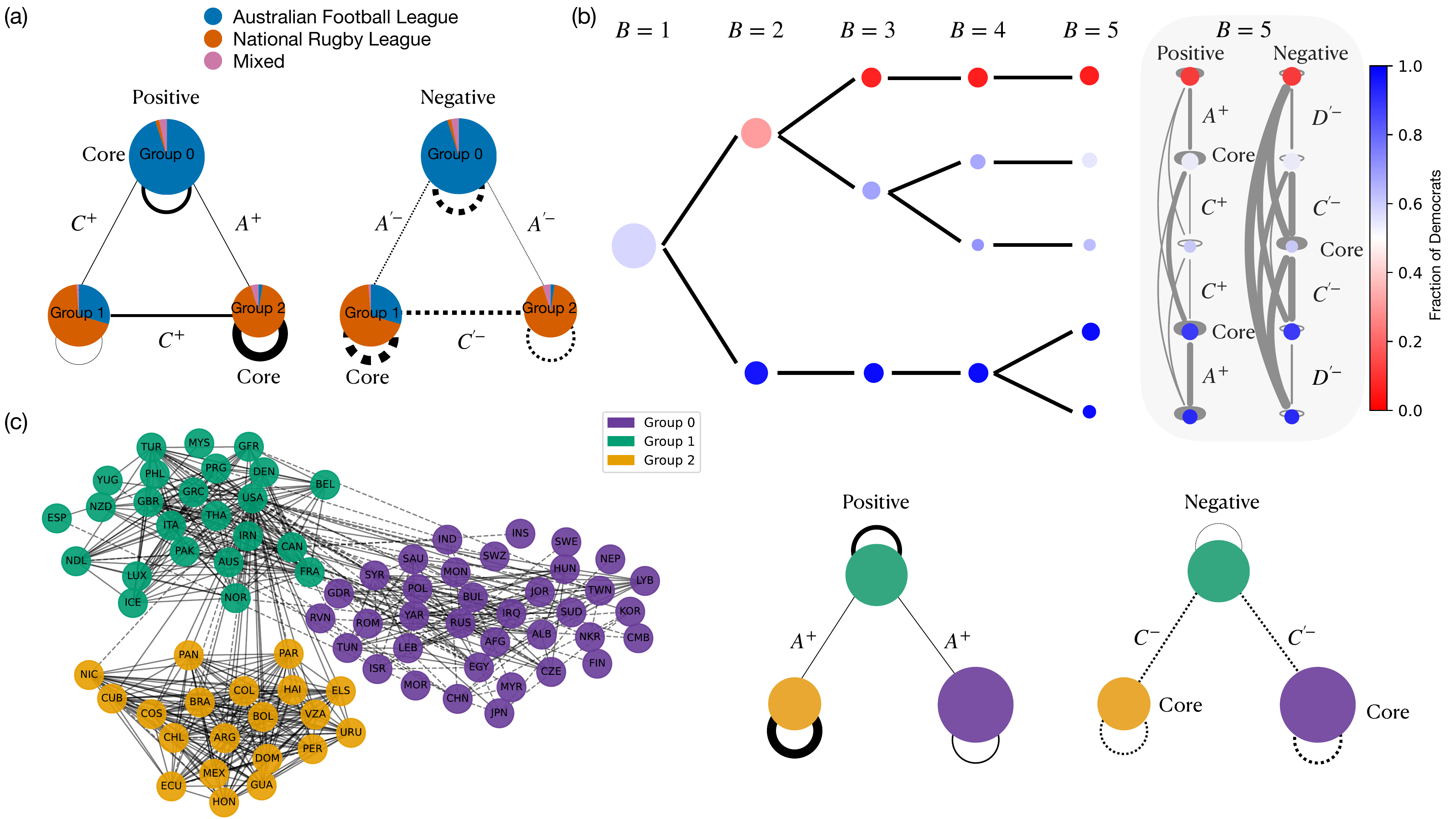}
\caption{\textbf{Examples of core-periphery structures in signed networks.} (a) Twitter signed network~\cite{pougue2023learning}. The graphical representation of the edge density matrix corresponding to the three top-level groups identified by WSBM. Each circle (node) represents a community, and the colored sectors within each node illustrate the topic composition: Australian Football League (AFL), National Rugby League (NRL), and mixed-interest discussions.  
(b) USA Congress H100~\cite{neal2014backbone}. The partition into groups was obtained by the LSBM method with an increasing number of communities $B\in[0,5]$. The plot shows schematically how communities divide and the type of pairwise interaction they share. The diagram on the right displays the positive and negative density matrix for $B=5$. Colors indicate party affiliation of group members. (c) Correlates of Wars for the years 1954-57~\cite{doreian2015structural}. It shows the top-level network partitioned by WSBM (left) and the graphical representation of the edge density matrix for the communities (median and right).
The label on each node is the name of the region. In all plots, the size of each node reflects the size of the group, solid lines represent positive edges, and dashed lines represent negative edges.}
\label{fig:cases}
\end{figure}

Another frequent unbalanced structure in our survey of empirical networks in Sec.~\ref{sec:1} is the core-periphery structure. Traditionally, in such structures, the core represents a dense cluster of nodes with strong internal connectivity, while the periphery consists of loosely connected nodes that rely on the core for interaction~\cite{rombach2017core,betzel2018,liu2023nonassortative}.
For unsigned interactions in social media, the core typically comprises influencers or highly active users who act as information creators, whereas the periphery primarily consumes information~\cite{csermely2013structure,rombach2017core,bastos2018core}. 
While core–periphery structures have been extensively studied for positive or neutral interactions, their role in negative exchanges remains less clear. Our results reveal a distinct and previously under‑reported behaviour of the core under negative interactions, which we detail 
in Fig.~\ref{fig:cases} and below in three empirical examples:

\begin{itemize}
\item[(i)] A social-media network of 3,883 Twitter users discussing sports in Australia, with positive edges representing ``like" and negative edges representing opposing replies~\cite{pougue2023learning}. Our analysis of the network partition (Fig.~\ref{fig:cases}a) reveals two core-periphery structures in positive interactions and one core-periphery structure in negative interactions. 
By mapping the topical interests of users (metadata) to the inferred blocks, we identify that Groups 0 and 2 are each dominated by users mostly interested in different sports (Australian Football League and National Rugby League, respectively). As expected in these cases, the relationship between these two groups in the positive network is assortative, characterized by high internal positive edge densities. Conversely, Group 1 emerges as a mixed-topic hub. While it occupies a peripheral role in positive interactions, lacking a singular sports identity to drive cohesive `likes', it is a central core role in the negative interactions. This suggests that groups containing members of diverse interests or opinions bridge otherwise assortative topical clusters~\cite{bastos2018core}, serving as a critical structural bridge that sustains discourse across the entire network.
We note that a posterior-analysis of this network revealed plausible alternative partitions of the network, including types of mesoscale structures different from the one reported here (see SM-Table~\ref{tab:twitter}).

\item[(ii)] A well-known political network of the Congress of the United States of America (USA), constructed based on the co-sponsorship patterns of bills in the 100th House of Representatives~\cite{neal2014backbone}. Nodes are members of the House of Representatives, and edges represent significant positive or negative tendencies to co-sponsor bills.
Although our network partition (Fig.~\ref{fig:cases}b) was obtained without using party affiliation metadata, the inferred blocks align closely with partisan lines, providing a structural explanation for legislative behavior. 
While the balanced structure is dominant, we focus on the emergence of core–periphery structures that reveal more interesting patterns (confirmed by our posterior sampling analysis, see SM-Table~\ref{tab:congress}).
For positive interactions, the cores are composed of highly cohesive partisan blocks (e.g., relatively solid Democrat or Republican clusters) that exert influence over peripheral groups containing more mixed members of both parties. Most significantly, the negative interactions reveal a mixed-party core. While balance theory might predict that negative ties only exist between party blocks, we find that bipartisan clusters, containing both Democrats and Republicans, serve as the structural core of the negative interactions.

\item[(iii)] The well-studied Correlates of War network for the years 1954-57~\cite{doreian2015structural}. Nodes represent geopolitical regions and edges denote formal alliances $(+)$ or diplomatic/military conflicts $(-)$.
The partition (Fig.~\ref{fig:cases}c) reveals three main groups in assortative relationships in the positive interactions, but in core-periphery relationships in the negative interactions 
(the robustness of this pattern is confirmed by our posterior sampling analysis, see SM-Table~\ref{tab:CoW}). 
While Group 1 (comprising the USA and Western Europe) remains internally cohesive through a dense positive alliance, its broader global connectivity is primarily mediated through negative interactions. Specifically, its engagement with the rest of the world is defined by prominent negative edges with the conflictual cores. In this structure, Group 0, a geographically diverse cluster spanning Europe, Africa, and Asia, emerges as the primary negative core. This indicates that Group 0 functions as the central arena of global friction, connected with the established Western bloc (Group 1).
\end{itemize}

These examples reveal a common pattern: core-periphery relationships often appear in the negative interactions with negative cores often composed by diverse groups which accommodate internal conflicts (becoming the main arena for conflict), sit at the center of disputes, and connect different groups. In other words, they act as mixed hubs that concentrate antagonistic ties and mediate interactions across opposing factions. 
These structures are not accidental but functional. Ref~\cite{bastos2018core} points out that core-periphery formations are topologically optimized for the efficient diffusion of specialized information from an active core to a relatively passive periphery.
In the context of negative interactions in social networks, our results suggest that this specialized information often takes the form of controversial or polarizing discourse coming from groups with members with different affiliations.
Thus, these unbalanced structures act as structural bridges that concentrate disagreement at the center to maintain network-wide engagement, suggesting they are critical for understanding the dynamics of negotiation and disagreement in social and political networks~\cite{bastos2018core,askarisichani2019structural}.

\subsection{Micro and meso scales exhibit different balance}
\label{sec:micro_meso}

\begin{figure}[ht]
\centering
\includegraphics[width=1\linewidth]{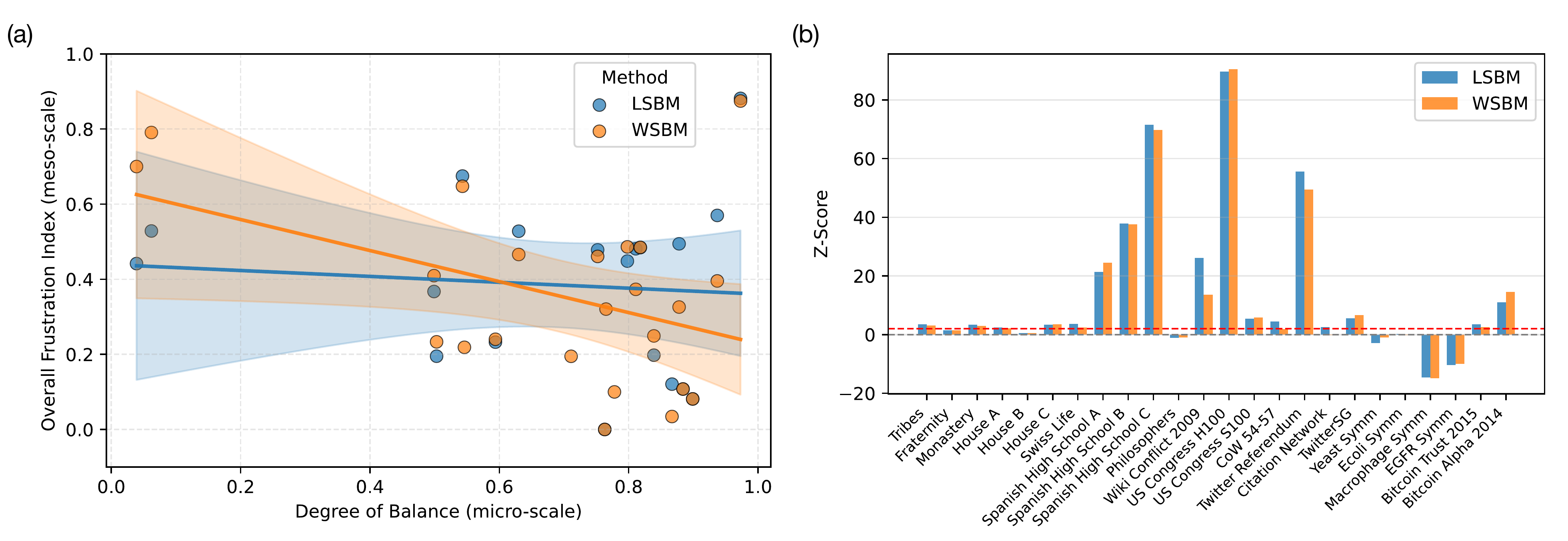}
\caption{\textbf{Microscale balance vs. mesoscale frustration.} (a) Negative relationship between DoB (fraction of balanced triads) and overall frustration index. 
Scatter plots show data points of each network with linear regression fits and 95\% confidence intervals (shaded regions).
(WSBM: $R^2=0.181, p=0.043$; LSBM: $R^2=0.008, p=0.708$).
(b) Z-scores quantify the deviation of observed DoB from null models preserving mesoscale block structure.
The red dashed line is when $z\text{-score} =2$.
Even when mesoscale structure is fixed, microscale balance is significantly overrepresented in real networks, indicating that micro-level balance and meso-level balance are not interchangeable.
Other relationships between mesoscale (im)balance and microscale balance also support this result (SM-Fig.~\ref{fig:3}).
The computation of DoB and $z\text{-score}$ refers to Methods~\ref{Balanced triads}-\ref{null-model}.
}
\label{fig:micro_meso}
\end{figure}

Structural balance is a scale-dependent property~\cite{kirkley2019balance}. Although perfectly balanced or anti-balanced states are equivalent across micro and meso scales under strong balance theory (Fig.~\ref{fig:rela}), measuring imperfect balance reveals important differences.
We quantify balance at two levels: local triadic balance (DoB) at the microscale and overall frustration at the mesoscale.
Contrary to the expectation that local balance implies global balance, Fig.~\ref{fig:micro_meso}a shows only a weak correlation between them (i.e., a weak negative correlation between the microscale degree of balance and the mesoscale frustration). This result indicates that microscale and mesoscale balance are not interchangeable: local structural balance does not reliably predict global frustration.
To isolate the effect of local structure from this broader mesoscale organization, we introduce null models sampled from SBMs as a baseline for comparison (see Methods~\ref{null-model}).
These null models preserve the mesoscale block structures and edge sign densities, allowing us to compute the z-score for observed DoB.
As shown in Fig.~\ref{fig:micro_meso}b, most empirical networks have a significantly higher microscale balance than expected under the fixed mesoscale structures.
Overall, these results confirm that the micro-level and meso-level balance are distinct dimensions of signed network structure, with networks with similar triad balance exhibiting vastly different levels of mesoscale balance (overall frustration).

\section{Discussion}

Social balance theory has been a cornerstone of social network analysis for over half a century~\cite{heider1946attitudes, cartwright1956structural}, providing a foundational framework for the study of signed networks, where positive ties denote friendship or alliance and negative ties represent enmity or conflict.
Traditionally, quantitative studies of signed networks have focused on micro-level triad analysis (e.g., ``the friend of my friend is my friend" and ``the enemy of my enemy is my friend").
While this micro-level analysis has provided valuable insights, it has an emphasis on local configurations and overlooks the complexity of the connectivity present in empirical networks, which includes different mesoscale structures. 
To address this gap, we proposed and applied a methodology to characterize mesoscale structures in signed networks that does not a priori assume that structures are balanced, being thus suitable to evaluate the extent into which the observations are explained by structural balance theory and to reveal previously unidentified (unbalanced) structures shaping the network.

Applying this methodology to 24 empirical networks, including social, political, citation, financial, and biological networks, we find that unbalanced mesoscale structures are ubiquitous, even when substantial balance exists at the microscale of triangles.
In many empirical networks, particularly social and political ones, the dominant structure is not balanced but instead assortative in both positive and negative interactions. 
The case study of Spanish High School students illustrates such a pattern of social homophily in both positive and negative interactions~\cite{mark2010}: the frequency of interaction often outweighs the sign of the connection; individuals interact within their social clusters regardless of the nature of the tie.
In contrast, as confirmed by our study on synthetic networks  (Fig.~\ref{fig:syn}), methods that are constrained to detect balanced partitions fail to capture this (and other) unbalanced relationships even when they are the prevailing mesoscale structure of signed networks.
While the particular synthetic case discussed above focused on $(A^+,A^{(')-})$ structures, which can potentially be recovered applying classical community-detection methods (e.g., Louvain), ignoring edge signs, this is not the case for other unbalanced structures (see SM-Fig.~\ref{fig:S8} for a similar analysis focusing on synthetic graphs with unbalanced non-assortative structures). In particular, only $4$ of the $18$ pairwise structures we define here (in Fig.~\ref{fig:class2}) are detected using traditional heuristic methods: the two balanced ones -- $(A^+,D^-)$ and $(A^+,D^{'-})$ -- using methods based on structural balance (e.g., signed Louvain); and the two assortative ones -- $(A^+,A^-)$ and $(A^+,A^{'-})$ -- using community-detection methods (e.g., traditional Louvain). The remaining $14$ structure types are invisible for these methods, requiring the combination of the inferential and classification approaches proposed here to be detected and interpreted.  
The lack of detection of non-balanced and non-assortative structures by heuristic methods is not due to detectability limits, which are present also in signed graphs (as studied in Ref.~\cite{bt2h-b7kb}). Instead, it is due to the embedded assumptions of assortativity and balance in these methods, rendering them systematically blind to broader classes of mesoscale organizations.

Another important finding of our work is that core-periphery structures play an important role in signed social networks. This includes not only the expected cases of well-connected influencers who create content consumed by the periphery~\cite{csermely2013structure,rombach2017core}, but also a new pattern in which a conflict-tolerant or mixed-affiliation group constitutes the core of a negative core-periphery relationship. While distinct, this negative core-periphery structure is theoretically consistent with prior research suggesting that core-periphery formations are topologically optimized for the efficient diffusion of specialized information from an active core to a relatively passive periphery~\cite{bastos2018core}. In this context, we posit that polarized discourse and inter-group conflict serve as the specialized information, with the mixed-group core acting as the primary generator of high-engagement, controversial, or diverse content. Consequently, the distinction between core structures in positive versus negative interactions serves to demarcate divergent behavioral patterns among users.

Future research should question assumptions of balance and consider the presence of unbalanced mesoscale structures in network representations. As our findings suggest, we can reasonably expect that studies with signed networks in other areas will be similarly successful in identifying diverse structures and providing new insights not only on the role of specific nodes but also on the organization and function of mesoscale communities and the network as a whole.
In particular, the application of our results to online social-media debates has the potential to reveal mesoscale structures beyond opinion polarization between two groups.
These analyses may require a generalization of our methods to directed and weighted signed networks \cite{liu2023nonassortative,hao2025social} (i.e., where influence may not be reciprocal) with the usage of more general SBMs and moving beyond a simple pairwise classification of community types to capture more multifactional structures.

\section{Materials and Methods}

\subsection{Data preprocessing}
\label{data}

We consider 18 social-political networks, 4 gene-regulatory networks, and 2 financial networks. A basic description of these networks can be found in the Supplementary Materials (SM).
While some networks have directed (e.g., \textit{Spanish High School}, \textit{TwitterSG}) and weighted (e.g., \textit{Spanish High School}, \textit{Monastery}) edges, 
here networks were standardized by mapping the existing data to undirected (symmetrized) and unweighted (binarized) signed graphs with a single connected component. This was done in order to ensure a uniform comparison across all methods -- while the SBMs can handle weighted directed edges, Louvain and spin-glass cannot -- and to align with structural balance theory in its standard form, which prioritizes the tie sign consistency over strength (but see Ref.~\cite{hao2025social} for a possible generalization of balanced triads to directed networks). 
This was achieved by first considering only nodes in the largest connected component and then following
the rules proposed in Refs.~\cite{aref2019balance,gallo2024testing}: if two nodes share the same sign in both directions, the undirected connection preserves that sign; if the connection is non-reciprocal, we symmetrize by adopting the existing sign; if the signs are opposite, the undirected edge is negative.
For graphs with a weighted configuration, we binarized the weights by assigning $+1$ to each positive edge and $-1$ to each negative edge.

Formally, given a directed signed adjacency matrix $\mathbf{S} = (S_{ij}) \in \mathbb{R}^{N \times N}$ where $S_{ij} >0$ indicating a positive edge $i \mapsto j$, $S_{ij} <0$ a negative edge, and $S_{ij} =0$ no edge, we construct a simple undirected signed adjacency matrix $\mathbf{A} = (A_{ij}) \in \mathbb{R}^{N \times N}$, using the symmetrization rule:
\begin{equation}
A_{ij} =
\begin{cases}
0, & S_{ij} = 0 \text{ and } S_{ji} = 0, \\[6pt]
-1, & S_{ij} S_{ji} < 0, \\[6pt]
\operatorname{sgn}(S_{ij} + S_{ji}), & \text{otherwise},
\end{cases}
\end{equation}
where $\operatorname{sgn}(x) = +1$ if $x > 0$, and $\operatorname{sgn}(x) = -1$ if $x < 0$.

Before pre-processing, we analyzed the prevalence of reciprocated edges and conflicting edges (i.e., directed edges between the same two nodes with opposing signs) among the directed networks we collected, see details for each network in SM-Table~\ref{tab:directed}. Conflicting edges are extremely rare across all datasets: 50\% of the directed networks have no conflicting edges, and the other 50\% have less than 5\% conflicting edges.
After this pre-processing, we calculated the descriptive statistics for the largest connected component of each network. The results are summarized in Table~\ref{tab:descriptive}.

\subsection{Frustration index}
\label{FI}
To measure the structural (im)balance on the mesoscopic scale, we use the frustration index~\cite{aref2019balance,gallo2024assessing} calculated based on the fraction of misplaced edges according to classical balance theory, i.e., the positive edges between communities, and the negative edges within communities.
We use two frustration indices in the main text. The first calculates the pairwise imbalance between communities, and the second calculates the overall imbalance of the partition.
Given a pair of communities $i$ and $j$, the pairwise frustration index is computed as
\begin{equation}
\text{FI}_{ij} = \frac{L^+_{ij}+L^-_{ii}+L^-_{jj}}{L_{ii}+L_{jj}+L_{ij}},
\end{equation}
where $L^+_{ij}$ represents the number of positive edges between community $i$ and $j$, $L^-_{ii}$ represents the number of negative edges within community $i$, $L_{ij}$ represents the total number of edges between community $i$ and community $j$, and $L_{ii}$ represents the total number of edges within the community $i$.

Given a partition with $B$ communities, the overall frustration index is defined as \cite{aref2019balance,gallo2024assessing}:
\begin{equation}
\text{FI} = \frac{\sum_{i=1}^{B} L^{-}_{ii} + \sum_{i=1}^{B} \sum_{j(>i)} L^+_{ij}}{L},
\end{equation}
where $\sum_{i=1}^{B} L^{-}_{ii}$ represents the total number of negative edges within each community, $\sum_{i=1}^{B} \sum_{j(>i)} L^+_{ij}$ the total number of positive edges between communities, and $L$ the total number of edges overall.
$\text{FI} \in [0,1]$, with $\text{FI}=1$ indicating complete anti-balance and $\text{FI}=0$ perfect balance.

\subsection{SBM inference}
\label{ssec:sbm}

Here we describe the statistical and computational procedures employed in the methods based on stochastic block models (SBM), including the two SBM approaches we employ in the survey (weighted and layered) and the corresponding hierarchical versions we used in case studies. The inference is performed using the Bayesian approach proposed in Refs.~\cite{peixoto2019bayesian,peixoto2014hierarchical} and implemented in Ref.~\cite{peixoto_graph-tool_2014}. 
In particular, the inference is non-parametric so that both the number of partitions $B$ and the allocation of nodes to them are obtained simultaneously, following the principle of minimum description length~\cite{peixoto_nonparametric_2017, peixoto2018nonparametric}. 
For each network, we obtain 10 independent Markov Chain Monte Carlo runs to characterize the posterior distribution of partitions and extract the representative partition that most samples are aligned with (see our code in Ref.~\cite{mesoscaleSG} for details). We perform our pairwise-classification of the relationship between groups using the consensus partition~\cite{peixoto2021revealing} obtained over these MCMC runs. By aggregating the outcomes over all plausible partitions, we focus on the most stable and statistically significant mesoscale organization of the network.

To evaluate the robustness of our mesoscale classification over alternative partitions, we quantify the variation of our classification over $1,000$ individual partitions sampled from the posterior distribution of the SBM model. These partitions were obtained by sampling  $100$ uncorrelated partitions (after equilibration) in each of the $10$ independent MCMC runs.
For each sampled partition, we extracted the dominant mesoscale type across all block pairs, recorded it as the sample winner, and compared it to the dominant mesoscale type obtained in the consensus partition. The certainty of the inferred dominant type was then quantified as the proportion of samples in which the dominant type prevailed, computed as
\begin{equation}
P = \frac{1}{k}\sum_{1}^{k}\delta(T^{(k)},T^{*}),
\end{equation}
where $T^{(k)}$ is the dominant type extracted from the k-th posterior sample, $T^{*}$ is the winner across all $k$ samples, and $\delta(a,b)=1$ if $a=b$ and $\delta(a,b)=0$ if $a \neq b$. A high certainty, with agreement with the result in the consensus partition, indicates the posterior is concentrated around one structure, meaning the network has a robust and unambiguous community organization. A low certainty suggests the posterior is diffuse across multiple structures, reflecting an ambiguity in the network's organization.

To quantify the uncertainty in the number of blocks $B$ in the partition, we use the effective number of blocks $B_e$. Unlike the discrete count $B$, $B_e$ is a continuous value that weights groups by their relative size, making it more sensitive to posterior fluctuations and less prone to zero-variance artifacts in averages. It is defined using the Shannon entropy of the group distribution:
\begin{equation}
B_e = \exp \left( -\sum_{r=1}^{B} p_r \ln p_r \right),
\end{equation}
where the weight $p_r =N_r/N$ is the fraction of nodes in group $r$. Note that $B_e=B$ only when all groups are of equal size; otherwise, $B_e<B$.

The results obtained for all networks in our survey and the two SBM methods are shown in the SM-Tables~\ref{tab:LSBM}-\ref{tab:WSBM} and Figs.~\ref{fig:S5}-\ref{fig:S6}. The results obtained for the four networks used in our case studies are reported in the SM-Tables~\ref{tab:spanish}-\ref{tab:CoW}. This analysis complements our bootstrap analysis (see Methods~\ref{Robustness}), which validates the robustness of the block assignments themselves, while the posterior sampling described here specifically measures the stability of our classification over different (plausible) partitions of the network.

\subsection{Classification robustness}
\label{Robustness}

Our community structure classification is computed in Eq.~(\ref{eq.adc}), for a given network partition and using the density matrices $w^\pm$ defined in Eq.~(\ref{eq.w}).
To evaluate the statistical confidence of this classification for a pair of groups $r$ and $s$, we adopt a bootstrapping method to estimate the robustness of the classification under small perturbations of the community structure while preserving the underlying density characteristics~\cite{liu2023nonassortative}. Given a signed graph $G$ (with $L^+$ positive and $L^-$ negative edges) and partition, we compute 
new positive and negative density matrices, $w'^{+}$ and $w'^{-}$, using a bootstrapping approach, i.e., 
randomly choosing $L^+$ and $L^-$ edges with replacement from the positive and negative edges (respectively) in $G$.
We repeat the process $k$ times and compute the certainty $R_{rs}$ for each pair of communities $r$ and $s$, as the fraction of all $k$ classifications $M'_{rs}$ that are equal to the original type $M_{rs}$ as
\begin{equation}
R_{rs}=\frac{1}{k}\sum_{1}^{k}\delta(M_{rs},M'_{rs}),
\end{equation}
where $M_{rs}$ is the pairwise classification obtained from Eq.~(\ref{eq.adc}), and 
$\delta(a,b)=1$ if $a=b$ and $\delta(a,b)=0$ if $a \neq b$.
This classification robustness score is the certainty for each pair of communities $r$ and $s$, calculated as the fraction of trials that maintain the original classification.

\subsection{NMI}
\label{NMI}
The normalized mutual information (NMI) is a widely used metric to measure the similarity of detected communities~\cite{ana2003robust,felippe2024network}.
Given a network with $N$ nodes, the partition $p= \{p_1,...,p_B \}$ with $B$ communities, and the ground truth $u = \{u_1, ..., u_A\}$ with $A$ groups, we calculate the NMI as:

\begin{equation}
\mathrm{NMI} =
\frac{
    2 \displaystyle\sum_{i=1}^{A} \sum_{j=1}^{B}
    \frac{|p_i \cap u_j|}{N}
    \log \left(
        \frac{\frac{|p_i \cap u_j|}{N}}{\frac{|p_i|}{N} \cdot \frac{|u_j|}{N}}
    \right)
}{
    - \displaystyle\sum_{i=1}^{A} \frac{|p_i|}{N} \log \frac{|p_i|}{N}
    - \displaystyle\sum_{j=1}^{B} \frac{|u_j|}{N} \log \frac{|u_j|}{N}
}
,
\end{equation}
which corresponds to the mutual information between $p$ and $u$, normalized by the combined entropy. $\text{NMI} \in [0,1]$, with $\text{NMI}=1$ indicating perfect alignment and $\text{NMI}=0$ no alignment (zero mutual information).

\subsection{Model selection}
\label{model selection}
Here we describe a Minimum Description Length (MDL) model selection framework that quantifies the information gained from combining edge signs in the SBM description of the signed network. We compare our signed SBM model against an alternative null model in which the network's mesoscale connectivity and the edge signs are strictly independent.
Rather than assuming signs universally improve clustering, our MDL framework explicitly measures whether the empirical sign distribution provides a compressible, structurally aligned signal.
First, we fit the signed model using WSBM by passing edge signs as a discrete-binomial covariate (assuming communities dictate both topology and signs) and extract its entropy $\Sigma_{\text{signed}}$~\cite{peixoto2018nonparametric}. The null model's description length is
\begin{equation}
\Sigma_{\text{null}} = S(A',b) + S(W),
\end{equation}
where $S(A',b)$ represents the entropy (compression) of the network using a standard, unsigned SBM and $S(W)$ is the description length of the signs. In the computation of $S(W)$ we consider the signs to be independent, identically distributed random variables governed solely by the network's overall sign imbalance, irrespective of community structure. This is analytically defined as the negative log-marginal likelihood of the network's signs 
\begin{equation}\label{eq.sw}
S(W) = -\ln P(W) = \ln \binom{L}{L^+} + \ln(L + 1),    
\end{equation}
where $L = L^+ + L^-$ is the total number of edges and $L^+$ is the number of positive edges (see SM Sec.~\ref{supp:derivation} for the derivation). 
Finally, by comparing these two models, any instance where our signed SBM yields a lower description length than the null hypothesis ($\Delta \Sigma = \Sigma_{\text{null}} - \Sigma_{\text{signed}} > 0$) provides evidence that the edge signs are not merely global noise, but are fundamentally informative of the network’s mesoscale structure.

\subsection{Balanced triads}
\label{Balanced triads}

In order to measure and compare the degree of balance (DoB)~\cite{cartwright1956structural,estrada2014walk,singh2017measuring,kirkley2019balance,talaga2023polarization} of the empirical networks here, we calculate the fraction of (weakly) balanced triads, which includes triads with all positive edges $T_{+++}$, all negative edges $T_{---}$, and two negative edges $T_{+--}$.
The fraction of triads is calculated by adjacency matrix $\mathbf{A} = \mathbf{A}^+ + \mathbf{A}^- $ ($\mathbf{A}^+ \equiv \{A_{ij}^+\}^N$, $\mathbf{A}^- \equiv \{A_{ij}^-\}^N$):
\begin{equation}
\begin{aligned}
T_{+++} &= \frac{\text{Tr}[(\mathbf{A}^+)^3]}{\text{Tr}[(\mathbf{A}^+ - \mathbf{A}^-)^3]},\\
T_{---} &= \frac{\text{Tr}[(-\mathbf{A}^-)^3]}{\text{Tr}[(\mathbf{A}^+ - \mathbf{A}^-)^3]},\\
T_{+--} &= \frac{3\text{Tr}[\mathbf{A}^+(-\mathbf{A}^-)^2]}{\text{Tr}[(\mathbf{A}^+ - \mathbf{A}^-)^3]},\\
T_{++-} &= \frac{3\text{Tr}[(\mathbf{A}^+)^2(-\mathbf{A}^-)]}{\text{Tr}[(\mathbf{A}^+ - \mathbf{A}^-)^3]}.
\end{aligned}
\end{equation}
The degree of balance (DoB) is then defined as the fraction of balanced triads $\text{DoB} = T_{+++} + T_{+--}$, and the weak degree of balance (WDoB) is then $\text{WDoB} = T_{+++} + T_{+--} + T_{---}$.

\subsection{Null-model testing}
\label{null-model}
To determine if observed network patterns are statistically significant or simply a byproduct of indirect network constraints, we employ two distinct null-model testing frameworks.

For our main analysis in Sec.~\ref{sec:micro_meso}, we utilize a mesoscale null model based on the Stochastic Block Model. This model allows us to isolate the effects of local structure from broader mesoscale organization. By sampling random networks from the inferred SBM (using either WSBM or LSBM), we generate a baseline that preserves the underlying block structure and the specific edge-sign densities within and between blocks.
This approach ensures that the observed patterns in our results are not merely a reflection of the network's community organization but are driven by specific social dynamics. The z-score derived from this model suggests the primary findings presented in this study.

For the characterization of empirical networks (summarized in Table~\ref{tab:descriptive}), we use a sign-randomization null-model. Since networks often contain an unequal fraction of positive and negative edges, raw counts alone are insufficient to quantify balance \cite{szell2010multirelational, hao2024proper}. Following Ref. \cite{kirkley2019balance}, we create a control ensemble by fixing the underlying network topology and randomly shuffling the edge signs. This preserves the global sign density while removing any specific structural correlations.

To quantify the deviations of our observed patterns from the null-model expectations, we calculate the z-score \cite{gallo2024testing}:
\begin{equation}
z(m) = \frac{N(m) - \langle N(m) \rangle}{\sigma(m)},
\end{equation}
where $N(m)$ is the number of pattern $m$ (different kinds of triads) appears in the empirical network, $\langle N(m) \rangle$ is the expected number of the same pattern under the null model, and $\sigma(m) = \sqrt{\langle N^2(m) \rangle - \langle N(m) \rangle^2}$ is the standard deviation of $N(m)$ under the null model. We adopt the convention where
$z(m) > 2$ indicates that the pattern $m$ is overrepresented, meaning it occurs significantly more than in the null model, and 
$z(m) < -2$ indicates that the pattern $m$ is underrepresented, meaning it occurs significantly less than in the null model.

\printbibliography[
    title={References},
    section=0
]

\section*{Acknowledgments}
We thank Fakhteh Ghanbarnejad for useful discussions and the anonymous reviewers for valuable comments that have improved this article.

\paragraph*{Funding:}
This project was funded by the Australian Research Council under project DP240100872, awarded to E.G.A, T.J.A., O.B., and T.P.P.

\paragraph*{Author contributions:}
Conceptualization: W.Z., E.G.A., T.J.A., and O.B.
Methodology: W.Z., T.P.P., and E.G.A.
Data curation: W.Z. and T.P.P.
Software: W.Z. and T.P.P.
Visualization: W.Z.
Formal analysis: W.Z., E.G.A, and T.P.P.
Writing – original draft: W.Z. and E.G.A.
Writing – review \& editing: All authors.
Supervision: E.G.A, T.J.A., and O.B.
Funding acquisition: E.G.A, T.J.A., O.B., and T.P.P.
Project administration: E.G.A.

\paragraph*{Competing interests:}
There are no competing interests to declare.

\paragraph*{Data, code, and materials availability:}
Data and code needed to evaluate and reproduce the results in the paper are publicly available in the repository  \url{https://doi.org/10.5281/zenodo.19412922}, except for the Swiss StudentLife data, which can be requested via \url{https://codebook.sn.ethz.ch/}.
This study did not generate new materials.

\subsection*{Supplementary Materials}
Supplementary Text S1 to S3\\
Figs. S1 to S8\\
Tables S1 to S8\\

\clearpage
\newpage

\renewcommand{\thefigure}{S\arabic{figure}}
\renewcommand{\thetable}{S\arabic{table}}
\renewcommand{\theequation}{S\arabic{equation}}
\renewcommand{\thepage}{S\arabic{page}}
\setcounter{figure}{0}
\setcounter{table}{0}
\setcounter{equation}{0}
\setcounter{page}{1}


\begin{center}
\section*{Supplementary Materials for\\ Mesoscale structures in signed networks}

\vspace{20pt}
Wei Zhang$^{\ast}$,
Olga Boichak,
Tristram J. Alexander,\\
Tiago P. Peixoto,
Eduardo G. Altmann
\\ 
\small$^\ast$Corresponding author. Email: wei.zhang4@sydney.edu.au \\
\end{center}

\vspace{30pt}

\subsubsection*{This PDF file includes:}
Supplementary Text S1 to S3\\
Figures S1 to S8\\
Tables S1 to S8\\

\newpage

\setcounter{section}{0}
\renewcommand{\thesection}{S\arabic{section}}

\section{Data description}

We considered three kinds of datasets: social-political networks, financial networks, and gene-regulatory networks.
Although some networks originate from similar contexts, they vary substantially in structural characteristics such as size, edge density, and the ratio of negative to positive edges. These differences allow us to capture heterogeneity within the same domain and test the robustness of our method across diverse network configurations.

There are 18 social-political networks. 
\textit{New Guinea Tribes} \cite{read1954cultures} reflected political alliances and oppositions among 16 tribes, developed through fieldwork involving observations and conversations with leaders and members in the Central Highlands.
\textit{Fraternity} \cite{newcomb1961acquaintance} was created from a 15-week survey of 17 fraternity brothers at the University of Michigan in 1956, representing relationships with positive or negative edge values.
\textit{Monastery} \cite{sampson1968novitiate} was collected via a questionnaire asking each monk to list “3 brothers who you like the most and the least” among 18 monks at an American monastery.
\textit{College House A, B, C} \cite{lemann1952group} come from a small group survey on evaluations of other group members (sorority sisters) based on a range of behavioral characteristics.
\textit{Swiss Student} \cite{voros2021swiss} collected information about the relationships among Swiss students who shared most of their classes and often did coursework in groups.
\textit{Spanish High School A, B, C} \cite{ruiz2023triadic} gathers data on student relationships by having students name their connections and rate them from $-2$ to $+2$.
\textit{Philosophers} \cite{collins2009sociology} compiled master-pupil and acquaintanceship ties among philosophers from 800 B.C.E to 1935 C.E., based on historical texts in Randall Collins’ seminal book.
\textit{Wiki Conflict} \cite{brandes2009network} represents conflicts between English Wikipedia users, where nodes represent users and edges indicate positive or negative interactions. A negative interaction occurs when one user reverts another's edit.
\textit{U.S. Congress} \cite{neal2014backbone} illustrates bill co-sponsorship tendencies in the House and Senate. This signed network is derived from a bipartite network using the Stochastic Degree Sequence Model \cite{neal2014backbone}. Positive and negative edges indicate significant tendencies to co-sponsor or not.
\textit{Correlates of Wars} \cite{doreian2015structural} offers data on international political relationships from 1946 to 1997, updated every four years. Positive edges indicate alliances or political agreements, while negative edges signify enmities or disagreements.
\textit{Twitter Referendum} \cite{ordozgoiti2020finding} is an undirected signed network capturing Twitter user sentiments on the 2016 Italian Referendum, where edges are weighted $-1$ for opposing stances and $+1$ for the same stance.
\textit{Signed Citation network} \cite{kumar2016structure} consists of papers from the field of Computational Linguistics. The positive (endorsement), negative (criticism), and neutral citations are classified by keyword-based citation sentiment. Following \cite{song2022quantifying}, we combine neutral and positive citations as positive, since authors do not express clear opposing views in neutral citations.
\textit{TwitterSG}
\cite{pougue2023learning} is a signed network drawn from Twitter interactions between 20th May and 8th August 2021. Here, we select the popular topics of the Australian Football League and the National Rugby League. A positive connection exists from user A to user B if user A likes a tweet posted by user B. Conversely, a negative connection exists from user A to user B if user A expresses opposition to user B's tweet in a reply.

In gene-regulatory networks \cite{iacono2010determining}, \textit{Yeast} and \textit{Escherichia coli} are classified as transcriptional networks, while the networks involving the Epidermal Growth Factor Receptor (\textit{EGFR}) and \textit{Macrophage} are signaling networks. The \textit{Yeast} network relates to Saccharomyces cerevisiae, and the \textit{Escherichia coli} network pertains to Escherichia coli. \textit{Macrophage} are white blood cells that eliminate harmful substances like cancer cells, while \textit{EGFR} is linked to a protein that promotes cell division and survival in epidermal tissues.

The financial networks consist of Bitcoin traders, with \textit{Bitcoin OTC} and \textit{Bitcoin Alpha} \cite{kumar2016edge} built on user ratings of trust on trading platforms, ranging from $-10$ for fraudsters to $+10$ for complete trust.
The decision to symmetrize directed networks was primarily driven by the need for a fair, cross-method comparison, as baseline methods like Louvain and Spin-glass are designed for undirected graphs.

{\small
\begin{longtable}[c]{lccccccccc}
\multicolumn{10}{p{\textwidth}}%
{{\bfseries Table \thetable\ Dataset of directed empirical networks.}} \\
\multicolumn{10}{p{\textwidth}}%
{\small The table shows the total number of nodes $N$, the total number of edges $L$, the ratio of negative and positive edges $L^-/L^+$, and counts (with percentages) of +/– unidirectional, +/– reciprocated, conflicting links, and self loops. The percentages represent the relative frequency of each link type as a proportion of the total number of links in the network.
} \\
\label{tab:directed} \\
\hline
Name & $N$ & $L$ & $\frac{L^-}{L^+}$ & \makecell{$\xrightarrow{+}$\\ (\%)} & \makecell{$\xrightarrow{-}$\\ (\%)} & \makecell{$\overset{+}{\leftrightarrow}$ \\ (\%)} & \makecell{$\overset{-}{\leftrightarrow}$ \\ (\%)} & \makecell{conflicting \\ (\%)} & \makecell{self loop \\ (\%)} \\ \hline
\endfirsthead
\multicolumn{10}{l}%
{{\bfseries Table \thetable\ continued from previous page}} \\
\hline
Name & $N$ & $L$ & $\frac{L^-}{L^+}$ & \makecell{$\xrightarrow{+}$\\ (\%)} & \makecell{$\xrightarrow{-}$\\ (\%)} & \makecell{$\overset{+}{\leftrightarrow}$ \\ (\%)} & \makecell{$\overset{-}{\leftrightarrow}$ \\ (\%)} & \makecell{conflicting \\ (\%)} & \makecell{self loop \\ (\%)} \\ \hline
\endhead
\hline
\endfoot
\endlastfoot
Monastery \cite{sampson1968novitiate} & 18 & 103 & 0.84 & \makecell{25\\ (24\%)} & \makecell{28\\ (27\%)} & \makecell{30\\ (29\%)} & \makecell{18\\ (17\%)} & \makecell{2\\ (2\%)} &\makecell{0\\ (0\%)} \\
College House A \cite{lemann1952group} & 21 & 94 & 0.84 &  \makecell{24\\ (26\%)} & \makecell{36\\ (38\%)} & \makecell{26\\ (28\%)} & \makecell{6\\ (6\%)} & \makecell{2\\ (2\%)} &\makecell{0\\ (0\%)} \\
College House B \cite{lemann1952group} & 17 & 83 & 1.02 & \makecell{15\\ (18\%)} & \makecell{30\\ (36\%)} & \makecell{24\\ (29\%)} & \makecell{10\\ (12\%)} & \makecell{4\\ (5\%)} &\makecell{0\\ (0\%)} \\
College House C \cite{lemann1952group} & 20 & 81 & 1 & \makecell{22\\ (28\%)} & \makecell{26\\ (32\%)} & \makecell{18\\ (22\%)} & \makecell{14\\ (18\%)} & \makecell{0\\ (0\%)} &\makecell{1\\ (1\%)} \\
Swiss Studentlife \cite{voros2021swiss} & 181 & 713 & 0.21 &\makecell{278\\ (39\%)} & \makecell{108\\ (15\%)} & \makecell{307\\ (43\%)} & \makecell{10\\ (1\%)} & \makecell{10\\ (1\%)} &\makecell{0\\ (0\%)} \\
Spanish High School A \cite{ruiz2023triadic} & 409 & 8,557 & 0.17 &\makecell{3,425\\ (40\%)} & \makecell{1,036\\ (12\%)} & \makecell{3,780\\ (44\%)} & \makecell{122\\ (1\%)} & \makecell{194\\ (2\%)} &\makecell{0\\ (0\%)} \\
Spanish High School B \cite{ruiz2023triadic} & 238 & 3,755 & 0.11 &\makecell{1,755\\ (47\%)} & \makecell{268\\ (7\%)} & \makecell{1,580\\ (42\%)} & \makecell{28\\ (1\%)} & \makecell{124\\ (3\%)} &\makecell{0\\ (0\%)} \\
Spanish High School C \cite{ruiz2023triadic} & 534 & 12,812 & 0.16 &\makecell{4,843\\ (38\%)} & \makecell{1,399\\ (11\%)} & \makecell{5,992\\ (47\%)} & \makecell{200\\ (2\%)} & \makecell{378\\ (3\%)} &\makecell{0\\ (0\%)} \\
Citations \cite{kumar2016structure} & 8,711 & 25,161 & 0.09  &\makecell{22,910\\ (91\%)} & \makecell{2,157\\ (9\%)} & \makecell{57\\ (0\%)} & \makecell{0\\ (0\%)} & \makecell{14\\ (0\%)} &\makecell{23\\ (0\%)} \\
TwitterSG \cite{pougue2023learning} & 3,883 & 6,338 & 0.91 &\makecell{3,290\\ (52\%)} & \makecell{2,832\\ (45\%)} & \makecell{17\\ (0\%)} & \makecell{182\\ (3\%)} & \makecell{17\\ (0\%)} &\makecell{0\\ (0\%)} \\
Yeast \cite{iacono2010determining} & 664 & 1,064 & 0.26 &\makecell{844\\ (79\%)} & \makecell{220\\ (21\%)} & \makecell{0\\ (0\%)} & \makecell{0\\ (0\%)} & \makecell{0\\ (0\%)} &\makecell{0\\ (0\%)} \\
Escherichia coli \cite{iacono2010determining} & 1,376 & 3,150 & 0.70 &\makecell{931\\ (67\%)} & \makecell{466\\ (33\%)} & \makecell{0\\ (0\%)} & \makecell{0\\ (0\%)} & \makecell{0\\ (0\%)} &\makecell{0\\ (0\%)} \\
EGFR \cite{iacono2010determining} & 313 & 755 & 0.50 &\makecell{499\\ (66\%)} &\makecell{256\\ (34\%)} & \makecell{220\\ (21\%)} & \makecell{0\\ (0\%)} & \makecell{0\\ (0\%)} & \makecell{0\\ (0\%)} \\
Macrophage \cite{iacono2010determining} & 660 & 1,397 & 0.50 &\makecell{931\\ (67\%)} & \makecell{466\\ (33\%)} & \makecell{0\\ (0\%)} & \makecell{0\\ (0\%)} & \makecell{0\\ (0\%)} & \makecell{0\\ (0\%)} \\
Bitcoin OTC, 2015 \cite{kumar2016edge} & 348 & 998 & 0.09 &\makecell{269\\ (27\%)} & \makecell{63\\ (6\%)} & \makecell{646\\ (65\%)} & \makecell{12\\ (1\%)} & \makecell{8\\ (1\%)} &\makecell{0\\ (0\%)} \\
Bitcoin Alpha, 2014 \cite{kumar2016edge} & 687 & 2,714 & 0.18 &\makecell{660\\ (24\%)} & \makecell{332\\ (12\%)} & \makecell{1,618\\ (60\%)} & \makecell{62\\ (2\%)} & \makecell{42\\ (2\%)} &\makecell{0\\ (0\%)} \\
\hline
\end{longtable}
}

The statistics and basic description of the dataset after preprocessing are summed up in Table \ref{tab:descriptive}. Many of the empirical networks here are characterized by a small density $c=2L/N(N-1)$ and a small fraction of negative links $L^-/L^+$.

\begin{figure}[ht!]
\centering
\includegraphics[width=1\linewidth]{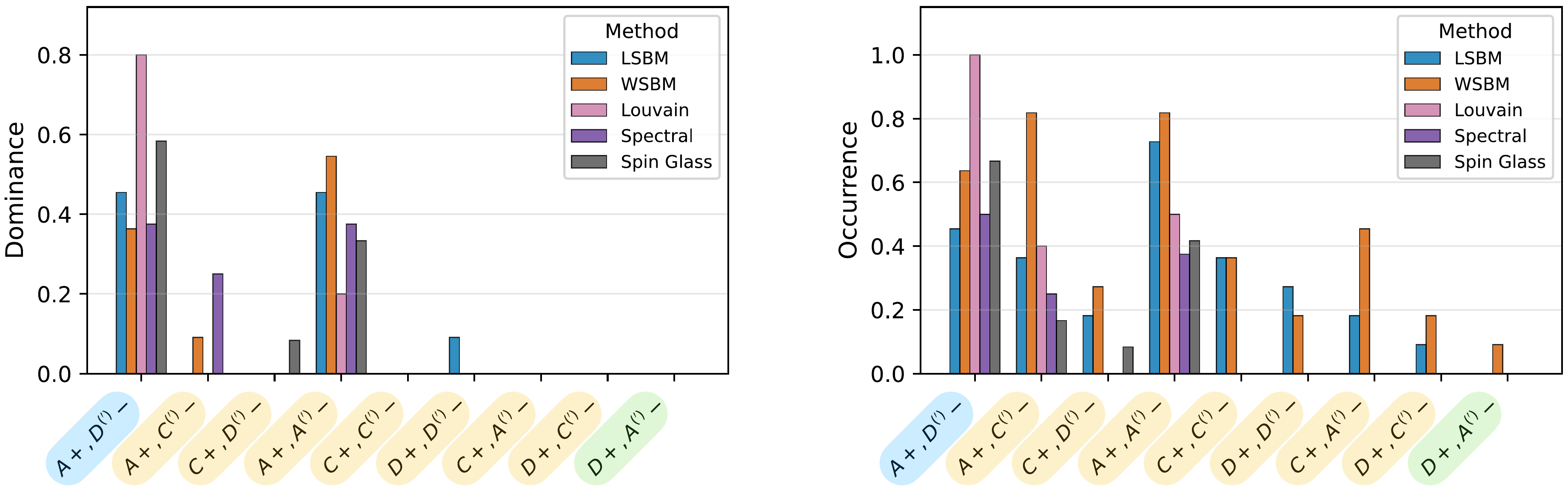}
\caption{\textbf{Community types in social-political networks.} Left: fraction of networks in which each community type is dominant. Right: fraction of networks in which each community type occurs.}
\label{fig:S1}
\end{figure}

\begin{figure}[ht!]
\centering
\includegraphics[width=1\linewidth]{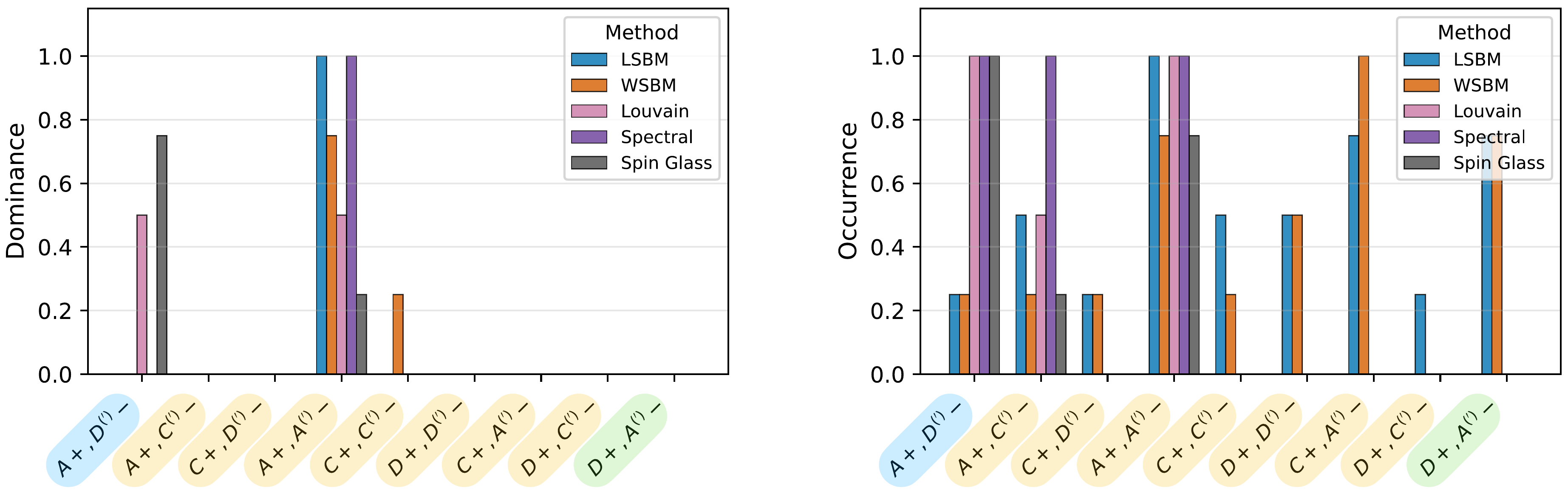}
\caption{\textbf{Community types in gene-regulatory networks.} Left: fraction of networks in which each community type is dominant. Right: fraction of networks in which each community type occurs.}
\label{fig:S2}
\end{figure}

\begin{figure}[ht!]
\centering
\includegraphics[width=1\linewidth]{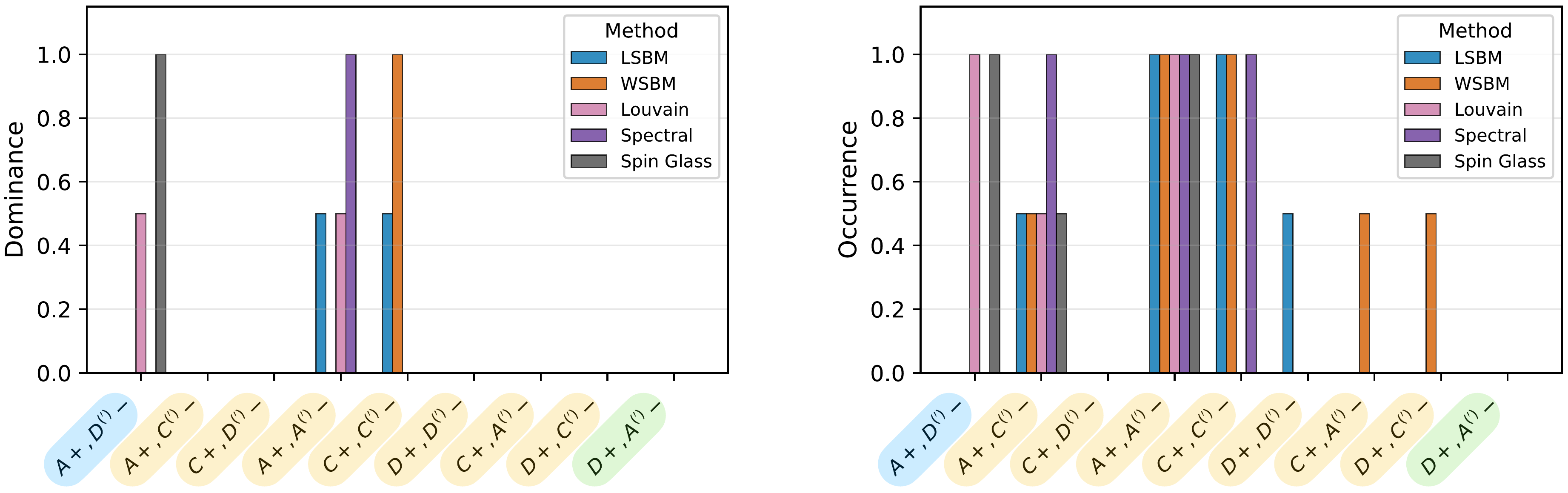}
\caption{\textbf{Community types in financial networks.} Left: fraction of networks in which each community type is dominant. Right: fraction of networks in which each community type occurs.}
\label{fig:S3}
\end{figure}

Our dataset consists primarily of social networks, reflecting the availability of signed network data in this domain. 
To assess whether this overrepresentation affects our findings, we examined within each domain separately (Figs.~\ref{fig:S1}-\ref{fig:S3}) and found that the result, that unbalanced structures are common, was qualitatively similar across domains. This suggests that the observed patterns are not driven solely by social networks but are consistent across different contexts.

{\small
\begin{longtable}[c]{lccccccccccc}
\multicolumn{12}{p{\textwidth}}%
{{\bfseries Table \thetable\ Descriptive statistics of the largest component of empirical networks.}} \\
\multicolumn{12}{p{\textwidth}}%
{\small The table shows the total number of nodes $N$, the total number of edges $L$, the ratio of negative and positive edges $L^-/L^+$, the average degree $\langle k \rangle$, the network density $c=2L/N(N-1)$, and  $T_{+++}$, $T_{++-}$, $T_{+--}$ and $T_{---}$ are the fractions of 3-cycles (triads) with all positive edges, one negative edge, two negative edges, and all negative edges, respectively. Their corresponding z-scores are shown in parentheses (see Methods~\ref{Balanced triads}). DoB is the fraction of balanced triads, that is, the fraction of triads with even negative edges.
Most empirical networks exhibit strong negative z-scores for triads with two positive edges. This indicates that unbalanced triads are underrepresented. These findings provide strong support for classical balance theory, as this ``forbidden" configuration is notably avoided.
For social-political networks, the results consistently show positive z-scores for most positive triads, but mixed z-scores for the ``enemy's enemy" triad ($+--$) and the ``all enemies" triad ($---$). This suggests that some networks strongly adhere to the balance principle, while others do not. 
For example, the \textit{Spanish High School A, B, C} shows a significant overrepresentation of the ($---$) configuration, which supports the weak balance theory.
Gene-regulatory networks tend to suppress structural balance, avoiding both clusters of positive relationships ($+++$) and ``enemy's enemy" configurations ($+--$). Besides, transcriptional networks (\textit{Yeast} and \textit{Escherichia coli}) generally avoid unbalanced $(++-)$ configurations to maintain hierarchical control, signaling networks (\textit{EGFR} and \textit{Macrophage}) appear to favor them.
Financial networks show weaker balance effects overall, suggesting different underlying mechanisms than pure social balance theory.
} \\
\label{tab:descriptive} \\
\hline
Name & $N$ & $L$ & $\frac{L^-}{L^+}$ & $\langle k \rangle$ & c & \makecell{$T_{+++}$\\ ($z$-score)} & \makecell{$T_{++-}$\\ ($z$-score)} & \makecell{$T_{+--}$\\ ($z$-score)} & \makecell{$T_{---}$ \\ ($z$-score)} & DoB \\ \hline
\endfirsthead
\multicolumn{12}{l}%
{{\bfseries Table \thetable\ continued from previous page}} \\
\hline
Name & $N$ & $L$ & $\frac{L^-}{L^+}$ & $\langle k \rangle$ & $c$ & \makecell{$T_{+++}$\\ ($z$-score)} & \makecell{$T_{++-}$\\ ($z$-score)} & \makecell{$T_{+--}$\\ ($z$-score)} & \makecell{$T_{---}$ \\ ($z$-score)} & DoB \\ \hline
\endhead
\hline
\endfoot
\endlastfoot
\makecell[l]{New Guinea \\Tribes \cite{read1954cultures}} & 16 & 58 & 1 & 7.25 & 0.48 & \makecell{0.279\\ (4.48)} & \makecell{0.029 \\(-6.28)} & \makecell{0.588\\ (3.70)} & \makecell{0.103\\ (-0.42)} & 0.867\\
Fraternity \cite{newcomb1961acquaintance} & 17 & 40 & 0.74 & 4.71 & 0.29 & \makecell{0.556\\ (4.51)} & \makecell{0\\ (-3.82)} & \makecell{0.222\\ (-0.89)} & \makecell{0.222\\ (2.73)} & 0.778\\
Monastery \cite{sampson1968novitiate} & 18 & 78 & 0.95 & 8.67 & 0.51 & \makecell{0.182\\ (1.89)} & \makecell{0.122\\ (-5.88)} & \makecell{0.583\\ (4.87)} & \makecell{0.113\\ (0.025)} & 0.765\\
\makecell[l]{College House \\ A \cite{lemann1952group}} & 21 & 77 & 1.08 & 7.33 & 0.37 & \makecell{0.229\\ (4.08)} & \makecell{0.133\\ (-4.54)} & \makecell{0.482\\ (1.77)} & \makecell{0.157\\ (0.60)} & 0.711\\
\makecell[l]{College House \\ B \cite{lemann1952group}} & 17 & 64 & 1.37 & 7.53 & 0.47 & \makecell{0.065\\ (-0.20)} & \makecell{0.221\\ (-1.85)} & \makecell{0.481\\ (0.91)} & \makecell{0.234\\ (1.23)} & 0.546\\
\makecell[l]{College House \\ C \cite{lemann1952group}} & 20 & 65 & 1.03 & 6.50 & 0.34 & \makecell{0.211\\ (2.0)} & \makecell{0.038\\ (-4.93)} & \makecell{0.673\\ (4.63)} & \makecell{0.077\\ (-1.27)} & 0.884\\
\makecell[l]{Swiss \\ Studentlife \cite{voros2021swiss}} & 181 & 546 & 0.28 & 6.03 & 0.034 & \makecell{0.689\\ (6.73)} & \makecell{0.15\\ (-9.15)} & \makecell{0.15\\ (2.22)} & \makecell{0.011\\ (0.26)} & 0.839\\
\makecell[l]{Spanish High \\School A \cite{ruiz2023triadic}} & 409 & 6,509 & 0.22 & 31.83 & 0.078 & \makecell{0.629\\ (16.36)} & \makecell{0.182\\ (-51.20)} & \makecell{0.169\\ (37.41)} & \makecell{0.020\\ (29.89)} & 0.798\\
\makecell[l]{Spanish High \\School B \cite{ruiz2023triadic}} & 238 & 2,889 & 0.14 & 24.28 & 0.102 & \makecell{0.729\\ (5.84)} & \makecell{0.183\\ (-15.15)} & \makecell{0.082\\ (18.75)} & \makecell{0.005\\ (9.71)} & 0.811\\
\makecell[l]{Spanish High \\School C \cite{ruiz2023triadic}} & 534 & 9,527 & 0.22 & 35.68 & 0.033 & \makecell{0.614\\ (14.67)} & \makecell{0.231\\ (-46.91)} & \makecell{0.138\\ (36.54)} & \makecell{0.017\\ (34.37)} & 0.752\\
Philosophers \cite{collins2009sociology} & 218 & 259 & 0.126 & 2.38 & 0.011 & \makecell{0.75\\ (0.38)} & \makecell{0.083\\ (-1.42)} & \makecell{0.167\\ (2.49)} & \makecell{0\\ (-0.14)} & 0.917\\ 
\makecell[l]{Wiki Conflict, \\2009 \cite{brandes2009network}} & 359 & 2,535 & 1.33 & 14.12 & 0.039 & \makecell{0.088\\ (0.36)} & \makecell{0.109\\ (-5.46)} & \makecell{0.675\\ (11.53)} & \makecell{0.128\\ (-1.41)} & 0.763\\
\makecell[l]{U.S. Congress, \\H100 \cite{neal2014backbone}} & 446 & 50,688 & 2.64 & 227 & 0.51 & \makecell{0.102\\ (460.54)} & \makecell{0.018\\ (-210.88)} & \makecell{0.492\\ (141.77)} & \makecell{0.388\\ (5.82)} & 0.594\\
\makecell[l]{U.S. Congress, \\S100 \cite{neal2014backbone}} & 101 & 2,143 & 2.52 & 42.44 & 0.42 & \makecell{0.116\\ (68.99)} & \makecell{0.02\\ (-34.48)} & \makecell{0.387\\ (-13.05)} & \makecell{0.478\\ (17.19)} & 0.503\\
\makecell[l]{Correlates of \\ Wars 1954-57 \cite{doreian2015structural}} & 80 & 492 & 0.178 & 12.3 & 0.16 & \makecell{0.869\\ (14.06)} & \makecell{0.093\\ (-16.91)} & \makecell{0.03\\ (-3.84)} & \makecell{0.008\\ (3.81)} & 0.899\\
\makecell[l]{Twitter \\ Referendum \cite{ordozgoiti2020finding}} & 10,864 & 251,396 & 0.054 & 46.28 & 0.004 & \makecell{0.938\\ (48.21)} & \makecell{0.027\\ (-71.41)} & \makecell{0.035\\ (150.56)} & \makecell{0\\ (-15.91)} & 0.973\\
\makecell[l]{Citations \cite{kumar2016structure}} & 8,711 & 25,073 & 0.094 & 5.76 & 0.001 & \makecell{0.779\\ (1.32)} & \makecell{0.179\\ (-3.54)} & \makecell{0.039\\ (6.67)} & \makecell{0.003\\ (3.41)} & 0.818\\
\makecell[l]{TwitterSG \cite{pougue2023learning}} & 3,883 & 5,480 & 0.952 & 2.82 & 0.001 & \makecell{0.495 \\ (14.37)} & \makecell{0.292 \\ (-2.90)} & \makecell{0.135 \\ (-7.39)} & \makecell{0.078 \\ (-1.63)} & 0.63\\
Yeast \cite{iacono2010determining} & 664 & 1,064 & 0.26 & 3.20 & 0.005 & \makecell{0.486\\ (-0.13)} & \makecell{0.371\\ (-0.33)} & \makecell{0.057\\ (-0.95)} & \makecell{0.086\\ (6.91)} & 0.543\\
\makecell[l]{Escherichia \\ coli \cite{iacono2010determining}} & 1,376 & 3,150 & 0.705 & 4.58 & 0.003 & \makecell{0.251\\ (1.23)} & \makecell{0.374\\ (-2.49)} & \makecell{0.248\\ (-1.43)} & \makecell{0.127\\ (2.76)} & 0.499\\
EGFR \cite{iacono2010determining} & 313 & 755 & 0.51 & 4.82 & 0.015 & \makecell{0.028\\ (-10.42)} & \makecell{0.873\\ (17.09)} & \makecell{0.034\\ (-8.96)} & \makecell{0.064\\ (2.60)} & 0.062\\
Macrophage \cite{iacono2010determining} & 660 & 1,397 & 0.50 & 4.23 & 0.006 & \makecell{0.024\\ (-14.63)} & \makecell{0.806\\ (21.29)} & \makecell{0.015\\ (-12.62)} & \makecell{0.155\\ (16.55)} & 0.039\\
\makecell[l]{Bitcoin OTC,\\ 2015 \cite{kumar2016edge}} & 348 & 665 & 0.12 & 3.82 & 0.011 & \makecell{0.908\\ (4.69)} & \makecell{0.063\\ (-5.38)} & \makecell{0.029\\ (-0.19)} & \makecell{0\\ (-0.56)} & 0.937\\
\makecell[l]{Bitcoin Alpha,\\ 2014 \cite{kumar2016edge}} & 687 & 1,853 & 0.26 & 5.39 & 0.008 & \makecell{0.747\\ (9.61)} & \makecell{0.12\\ (-17.36)} & \makecell{0.131\\ (2.31)} & \makecell{0.002\\ (-2.71)} & 0.878\\
\hline
\end{longtable}
}

\begin{figure}[ht!]
\centering
\includegraphics[width=1\linewidth]{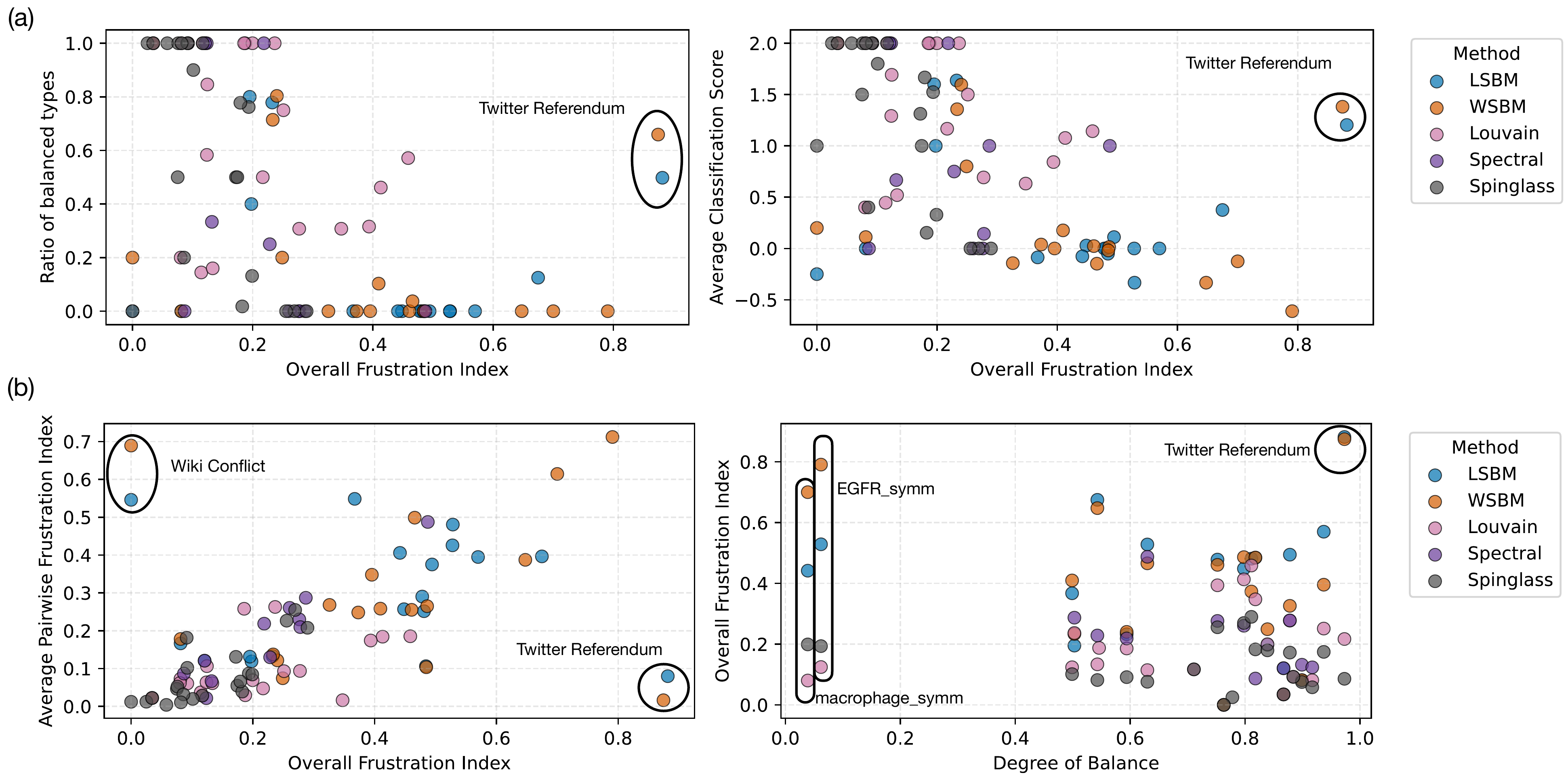}
\caption{
\textbf{Other relationships between mesoscale (im)balance and microscale balance. }
(a) Relationship between structure classifications and the overall FI given the partition of the network. 
Classification of balanced structure types (left panel), and average balance score (right panel), with the prediction from traditional balance theory, qualified by the overall frustration index. The results show that our classification closely matches theoretical structural balance expectations.
(b) Relationship between average pairwise FI, overall FI, and the ratio of balanced triads. 
The left panel confirms that the average pairwise frustration index could reflect the overall balance degree of the partition. In addition, the SBMs could find more unbalanced communities compared with other methods.
The right panel shows that the ratio of balanced triads from a microscopic perspective could not reflect the balance degree of a given partition at a mesoscopic level.
}
\label{fig:3}
\end{figure}

\newpage
\clearpage

\section{Implementation of SBMs}
\label{supp:SBM}

\begin{figure}[ht!]
\centering
\includegraphics[width=1\linewidth]{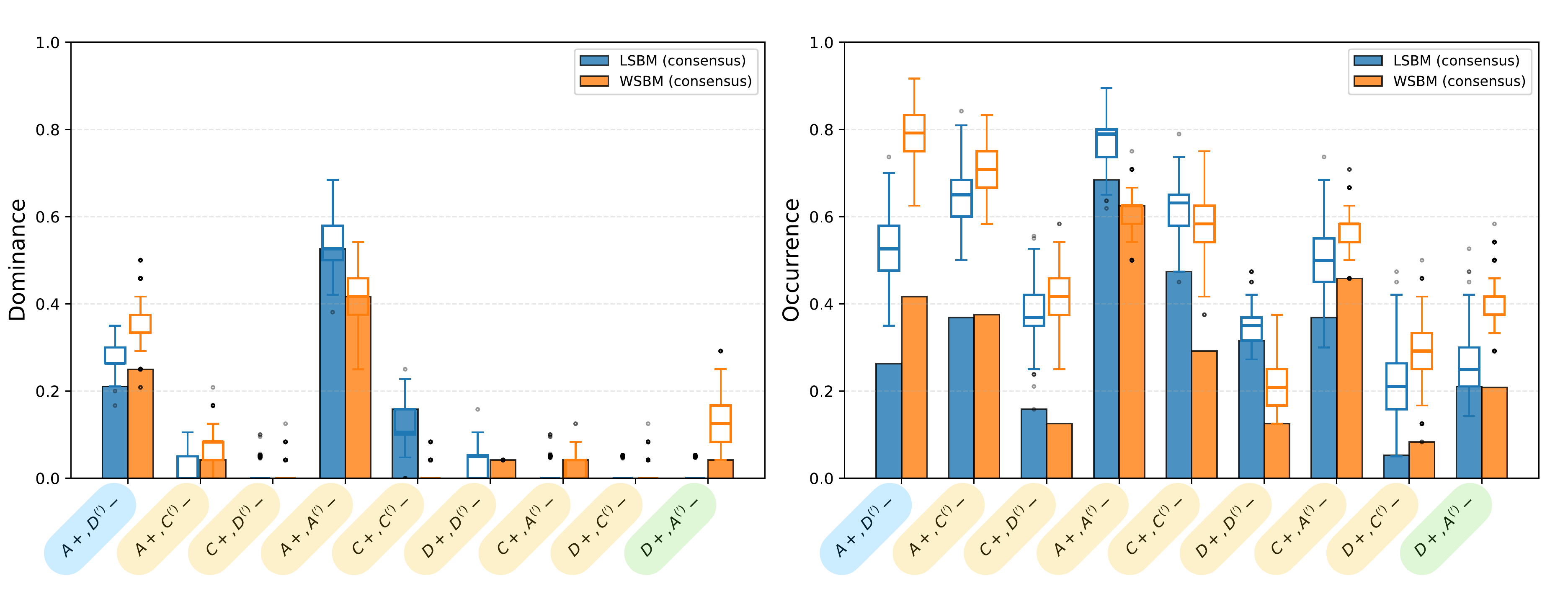}
\caption{\textbf{Dominance and occurrence of structure types under consensus and sampling.} 
Bar heights represent the consensus estimate: the fraction of networks where each type is dominant (left panel) or occurs (right panel), derived from the consensus partition. Overlaid box plots summarize the distribution of the same quantities across 1,000 sampling runs. Dominance is defined as the fraction of networks where the type is the most frequent; occurrence is defined as the fraction of networks where the type appears at least once. Results are shown for both LSBM (blue) and WSBM (orange). The red horizontal line within each box indicates the median across samples. The agreement between the bar height and the box plot location indicates robustness of the consensus estimate to sampling variability.}
\label{fig:S5}
\end{figure}

The comparison of consensus partition and the sampling from the posterior distribution are shown in the Table~\ref{tab:LSBM}, Table~\ref{tab:WSBM}, and Figure~\ref{fig:S5}. The posterior uncertainty analysis follows Methods~\ref{ssec:sbm}.
The posterior sampling initializes the state with a partition into one group. After we equilibrate the Markov chain following the principle of minimum description length, we collect partitions. The details implemented in Ref.~\cite{peixoto_graph-tool_2014}.
For most networks, there is almost the same between the classification obtained in the consensus partition and through the posterior sampling. The dominant type identified in the consensus partition remains the dominant type in the sampling analysis (i.e., 96\% consistency for LSBM and 92\% for WSBM). For the outliers, although the dominant type in the consensus differs from that in the sampling analysis, we found that this type is the second most common and occurs at a frequency very close to the dominant type.
Unbalanced structures were predominant, with 58\% of networks for LSBM and 67\% for WSBM. Specifically, the $(A^+,A^{(')-})$ configuration emerged as the most stable structure, particularly in large networks where the structural signal is strongest.

{\small
\begin{longtable}{
p{0.15\textwidth}   
p{0.1\textwidth}   
p{0.03\textwidth}   
p{0.03\textwidth}   
p{0.03\textwidth}   
p{0.02\textwidth}   
p{0.1\textwidth}   
p{0.03\textwidth}   
p{0.09\textwidth}   
p{0.09\textwidth}   
p{0.08\textwidth}   
}
\multicolumn{11}{p{\textwidth}}%
{{\bfseries Table \thetable\ Details of the inferred structures from posterior sampling by LSBM.}} \\
\multicolumn{11}{p{\textwidth}}%
{\small The table shows the consensus partition results (left) and posterior sampling analysis (right) by LSBM as a comparison. For the consensus results, we report (i) the dominant type among the retained blocks, (ii) the average robustness $\langle R \rangle$ of that dominant type (see Methods~\ref{Robustness}), (iii) the number of blocks $B$ after filtering out small communities (those with fewer than $\text{min}([0.05N,5])$ nodes are discarded), (iv) the effective number of blocks $B_e$,  and (v) the frustration index FI (see Methods~\ref{FI}).
For the posterior sampling (Methods~\ref{ssec:sbm}), we report: (i) the dominant type, representing the most frequent interaction type observed across the samples, (ii) the Certainty Score $P$, defined as the proportion of MCMC samples where the dominant type remained consistent, (iii) the average number of blocks ($B\pm \text{std}$) after filtering out small communities (those with fewer than $\text{min}([0.05N,5])$ nodes are discarded), (iv) the average effective number of blocks ($B_e\pm \text{std}$), and (v) the Frustration Index (FI $\pm \text{std}$) calculated over the samples.
In \textit{Bitcoin OTC 2015}, the consensus dominant type $(A^+,~A^{(')-})$ (marked by $^{\dagger}$) differs from the the most frequent interaction type  $(C^+,~C^{(')-})$ obtained in the sampling, but appears as the second most frequent type with a frequency close to the dominant one.
Because the posterior landscape is likely flat across multiple near-equivalent configurations.
} \\
\label{tab:LSBM} \\
\hline
Name 
& \multicolumn{5}{c}{Consensus} 
& \multicolumn{5}{c}{Sampling} \\ 
& Dominant & $\langle R \rangle$ & $B$ & $B_e$ & FI & Dominant & $P$ &  $\langle B \rangle$ & $B_e$ & FI  \\ 
\hline
\endfirsthead
\multicolumn{11}{l}%
{{\bfseries Table \thetable\ continued from previous page}} \\
\hline
Name 
& \multicolumn{5}{c}{Consensus} 
& \multicolumn{5}{c}{Sampling} \\ 
& Dominant & $\langle R \rangle$ & $B$ & $B_e$ & FI & Dominant & $P$ &  $\langle B \rangle$ & $B_e$ & FI  \\ 
\hline
\endhead
\hline
\endfoot
\endlastfoot
New Guinea Tribes     & $(A+,D^{(')}-)$ & 0.99 & 2 & 1.75 & 0.12 & $(A+,D^{(')}-)$ & 0.99 & 2.24 (45) & 2.03 (51) & 0.11 (5) \\
Fraternity            & - & - & 1 & 1.00 & - & - & 0.96 & 1.04 (18) & 1.03 (17) & - \\
Monastery             & - & - & 1 & 1.00 & - & - & 0.96 & 1.03 (19) & 1.03 (18) & - \\
College House A       & - & - & 1 & 1.00 & - & - & 0.63 & 1.45 (67) & 1.41 (62) & - \\
College House B       & - & - & 1 & 1.00 & - & - & 0.87 & 1.13 (35) & 1.11 (31) & - \\
College House C       & $(A+,D^{(')}-)$ & 0.66 & 2 & 1.84 & 0.11 & $(A+,D^{(')}-)$ & 0.98 & 1.99 (16) & 1.84 (14) & 0.11 (2) \\
Swiss Studentlife     & $(A+,A^{(')}-)$ & 0.78 & 8 & 6.97 & 0.27 & $(A+,A^{(')}-)$ & 0.99 & 6.96 (104) & 6.65 (96) & 0.28 (3) \\
Spanish High School A & $(A+,A^{(')}-)$ & 0.83 & 13 & 11.09 & 0.43 & $(A+,A^{(')}-)$ & 1.00 & 13.15 (107) & 11.30 (63) & 0.44 (1) \\
Spanish High School B & $(A+,A^{(')}-)$ & 0.78 & 11 & 10.30 & 0.38 & $(A+,A^{(')}-)$ & 1.00 & 11.75 (57) & 10.87 (38) & 0.43 (3) \\
Spanish High School C & $(A+,A^{(')}-)$ & 0.84 & 20 & 19.41 & 0.50 & $(A+,A^{(')}-)$ & 1.00 & 20.28 (45) & 19.75 (40) & 0.50 (0) \\
Philosophers          & - & - & 1 & 1.00 & - & - & 0.99 & 1.00 (0) & 1.00 (0) & - \\
Wiki Conflict 2009    & $(C+,C^{(')}-)$ & 0.83 & 4 & 3.03 & 0.00 & $(C+,C^{(')}-)$ & 0.86 & 3.90 (54) & 2.84 (46) & 0.00 (0) \\
US Congress H100      & $(A+,D^{(')}-)$ & 0.88 & 22 & 20.85 & 0.24 & $(A+,D^{(')}-)$ & 1.00 & 21.79 (61) & 20.63 (54) & 0.23 (0) \\
US Congress S100      & $(A+,D^{(')}-)$ & 0.84 & 6 & 5.50 & 0.17 & $(A+,D^{(')}-)$ & 1.00 & 6.13 (35) & 5.77 (40) & 0.19 (1) \\
CoW 1954-57           & $(A+,A^{(')}-)$ & 0.77 & 5 & 4.75 & 0.07 & $(A+,A^{(')}-)$ & 1.00 & 5.18 (39) & 4.89 (29) & 0.08 (4) \\
Twitter Referendum    & $(A+,D^{(')}-)$ & 0.91 & 64 & 32.44 & 0.90 & $(A+,D^{(')}-)$ & 1.00 & 63.18 (165) & 36.52 (123) & 0.88 (1) \\
Citations             & $(A+,A^{(')}-)$ & 0.91 & 41 & 30.25 & 0.61 & $(A+,A^{(')}-)$ & 1.00 & 41.52 (277) & 33.21 (295) & 0.62 (6) \\
Twitter SG            & $(A+,A^{(')}-)$ & 0.97 & 5 & 2.69 & 0.55 & $(A+,A^{(')}-)$ & 0.95 & 5.30 (46) & 3.13 (29) & 0.53 (2) \\
Yeast                 & $(D+,D^{(')}-)$ & 0.58 & 4 & 2.64 & 0.78 & $(D+,D^{(')}-)$ & 0.42 & 4.80 (87) & 3.13 (50) & 0.76 (1) \\
Escherichia coli      & $(A+,A^{(')}-)$ & 0.83 & 12 & 4.00 & 0.46 & $(A+,A^{(')}-)$ & 0.91 & 12.29 (140) & 6.56 (121) & 0.37 (8) \\
EGFR                  & $(A+,A^{(')}-)$ & 0.84 & 6 & 5.69 & 0.52 & $(A+,A^{(')}-)$ & 1.00 & 6.84 (78) & 6.15 (43) & 0.53 (2) \\
Macrophage            & $(A+,A^{(')}-)$ & 0.94 & 12 & 10.51 & 0.56 & $(A+,A^{(')}-)$ & 0.99 & 11.10 (104) & 9.89 (113) & 0.64 (12) \\
Bitcoin OTC 2015      & $(A+,A^{(')}-)$$^{\dagger}$ & 0.77 & 5 & 3.18 & 0.62 & $(C+,C^{(')}-)$ & 0.48 & 4.70 (69) & 3.40 (46) & 0.61 (3) \\
Bitcoin Alpha 2014    & $(C+,C^{(')}-)$ & 0.90 & 8 & 3.79 & 0.54 & $(C+,C^{(')}-)$ & 0.94 & 7.76 (63) & 4.34 (37) & 0.54 (2) \\
\hline
\end{longtable}
}

{\small
\begin{longtable}{
p{0.15\textwidth}   
p{0.1\textwidth}   
p{0.03\textwidth}   
p{0.03\textwidth}   
p{0.03\textwidth}   
p{0.02\textwidth}   
p{0.1\textwidth}   
p{0.03\textwidth}   
p{0.09\textwidth}   
p{0.09\textwidth}   
p{0.08\textwidth}   
}
\multicolumn{11}{p{\textwidth}}%
{{\bfseries Table \thetable\ Details of the inferred structures from posterior sampling by WSBM.}} \\
\multicolumn{11}{p{\textwidth}}%
{\small The table shows the consensus partition results (left) and posterior sampling analysis (right) by WSBM as a comparison. For the consensus results, we report (i) the dominant type among the retained blocks, (ii) the average robustness $\langle R \rangle$ of that dominant type (see Methods~\ref{Robustness}), (iii) the number of blocks $B$ after filtering out small communities (those with fewer than $\text{min}([0.05N,5])$ nodes are discarded), (iv) the effective number of blocks $B_e$, and (v) the frustration index FI (see Methods~\ref{FI}).
For the posterior sampling (Methods~\ref{ssec:sbm}), we report: (i) the dominant type, representing the most frequent interaction type observed across the samples, (ii) the Certainty Score $P$, defined as the proportion of MCMC samples where the dominant type remained consistent, (iii) the average number of blocks ($B\pm \text{std}$) after filtering out small communities (those with fewer than $\text{min}([0.05N,5])$ nodes are discarded), (iv) the average effective number of blocks ($B_e\pm \text{std}$), and (v) the Frustration Index (FI $\pm \text{std}$) calculated over the samples. 
For \textit{Monastery} and \textit{TwitterSG}, the consensus dominant type (marked by $^{\dagger}$) differs from the most frequent interaction type obtained via sampling. However, it is the second most frequent type and occurs at a frequency close to the sampling dominant type.
Because the posterior landscape is likely flat across multiple near-equivalent configurations.
} \\
\label{tab:WSBM} \\
\hline
Name 
& \multicolumn{5}{c}{Consensus} 
& \multicolumn{5}{c}{Sampling} \\ 
& Dominant & $\langle R \rangle$ & $B$ & $B_e$ & FI & Dominant & $P$ &  $\langle B \rangle$ & $B_e$ & FI  \\ 
\hline
\endfirsthead
\multicolumn{11}{l}%
{{\bfseries Table \thetable\ continued from previous page}} \\
\hline
Name 
& \multicolumn{5}{c}{Consensus} 
& \multicolumn{5}{c}{Sampling} \\ 
& Dominant & $\langle R \rangle$ & $B$ & $B_e$ & FI & Dominant & $P$ &  $\langle B \rangle$ & $B_e$ & FI  \\ 
\hline
\endhead
\hline
\endfoot
\endlastfoot
New Guinea Tribes       & $(A+,D^{(')}-)$ & 1.00 & 3  & 2.92 & 0.03 & $(A+,D^{(')}-)$ & 1.00 & 3.01(13)  & 2.93(12)  & 0.04(02) \\
Fraternity              & $(A+,C^{(')}-)$ & 0.76 & 2  & 1.73 & 0.10 & $(A+,C^{(')}-)$ & 0.97 & 2.01(10)  & 1.74(11)  & 0.10(03) \\
Monastery      & $(C+,C^{(')}-)$$^{\dagger}$ & 0.43 & 3  & 2.46 & 0.24 & $(A+,D^{(')}-)$ & 0.81 & 3.39(69)  & 3.22(63)  & 0.27(08) \\
College House A         & $(A+,D^{(')}-)$ & 0.67 & 4  & 3.21 & 0.19 & $(A+,D^{(')}-)$ & 0.94 & 3.63(61)  & 3.28(50)  & 0.25(07) \\
College House B         & $(A+,C^{(')}-)$ & 0.53 & 2  & 1.91 & 0.27 & $(A+,C^{(')}-)$ & 0.40 & 2.33(48)  & 2.22(49)  & 0.32(11) \\
College House C         & $(A+,D^{(')}-)$ & 0.66 & 2  & 1.84 & 0.11 & $(A+,D^{(')}-)$ & 1.00 & 2.03(16)  & 1.87(18)  & 0.11(02) \\
Swiss Studentlife       & $(A+,A^{(')}-)$ & 0.69 & 7  & 6.78 & 0.46 & $(A+,A^{(')}-)$ & 0.81 & 7.29(65)  & 6.92(67)  & 0.47(04) \\
Spanish High School 1   & $(A+,A^{(')}-)$ & 0.82 & 16 & 14.59& 0.48 & $(A+,A^{(')}-)$ & 1.00 & 18.14(85) & 16.02(67) & 0.53(02) \\
Spanish High School 2   & $(A+,A^{(')}-)$ & 0.73 & 12 & 11.17& 0.42 & $(A+,A^{(')}-)$ & 1.00 & 12.46(56) & 11.51(40) & 0.45(02) \\
Spanish High School 3   & $(A+,A^{(')}-)$ & 0.85 & 24 & 22.14& 0.48 & $(A+,A^{(')}-)$ & 1.00 & 24.75(68) & 22.69(69) & 0.49(01) \\
Philosophers            & $(D+,A^{(')}-)$ & 1.00 & 2  & 1.67 & 0.51 & $(D+,A^{(')}-)$ & 0.43 & 2.40(49)  & 2.17(39)  & 0.61(10) \\
Wiki Conflict 2009      & $(A+,A^{(')}-)$ & 0.96 & 7  & 4.95 & 0.00 & $(A+,A^{(')}-)$ & 1.00 & 6.62(50)  & 4.58(63)  & 0.00(00) \\
US Congress H100        & $(A+,D^{(')}-)$ & 0.93 & 30 & 28.54& 0.25 & $(A+,D^{(')}-)$ & 1.00 & 29.50(67) & 28.10(66) & 0.25(00) \\
US Congress S100        & $(A+,D^{(')}-)$ & 0.75 & 8  & 7.82 & 0.18 & $(A+,D^{(')}-)$ & 1.00 & 8.21(41)  & 8.03(39)  & 0.18(01) \\
CoW 1954-57             & $(A+,A^{(')}-)$ & 0.94 & 5  & 4.72 & 0.09 & $(A+,A^{(')}-)$ & 1.00 & 5.01(07)  & 4.74(06)  & 0.09(01) \\
Twitter Referendum      & $(A+,D^{(')}-)$ & 0.98 & 46 & 26.37& 0.88 & $(A+,D^{(')}-)$ & 1.00 & 47.60(191)& 30.06(163)& 0.88(00) \\
Citations               & $(A+,A^{(')}-)$ & 0.92 & 37 & 27.88& 0.56 & $(A+,A^{(')}-)$ & 1.00 & 37.00(184)& 29.33(194)& 0.61(03) \\
Twitter SG              & $(C+,A^{(')}-)$$^{\dagger}$ & 0.70 & 13 & 6.82 & 0.47 & $(A+,A^{(')}-)$ & 0.43 & 12.35(188)& 7.21(90)  & 0.49(03) \\
Yeast                   & $(A+,D^{(')}-)$ & 0.64 & 7  & 4.15 & 0.76 & $(A+,D^{(')}-)$ & 0.76 & 6.90(30)  & 4.36(24)  & 0.76(02) \\
Escherichia coli        & $(A+,A^{(')}-)$ & 0.75 & 16 & 7.72 & 0.30 & $(A+,A^{(')}-)$ & 0.95 & 17.84(142)& 9.52(53)  & 0.38(10) \\
EGFR                    & $(D+,A^{(')}-)$ & 0.90 & 7  & 6.74 & 0.92 & $(D+,A^{(')}-)$ & 0.95 & 7.99(149) & 7.47(154) & 0.92(02) \\
Macrophage              & $(A+,A^{(')}-)$ & 0.86 & 11 & 9.40 & 0.90 & $(A+,A^{(')}-)$ & 0.66 & 11.40(66) & 10.06(61) & 0.95(02) \\
Bitcoin OTC 2015        & $(D+,A^{(')}-)$ & 0.77 & 6  & 4.64 & 0.68 & $(D+,A^{(')}-)$ & 0.60 & 6.11(42)  & 4.86(46)  & 0.69(06) \\
Bitcoin Alpha 2014      & $(A+,A^{(')}-)$ & 0.91 & 12 & 7.03 & 0.54 & $(A+,A^{(')}-)$ & 0.81 & 11.44(115)& 7.96(75)  & 0.56(03) \\
\hline 
\end{longtable}
}

\begin{figure}[ht!]
\centering
\includegraphics[width=1\linewidth]{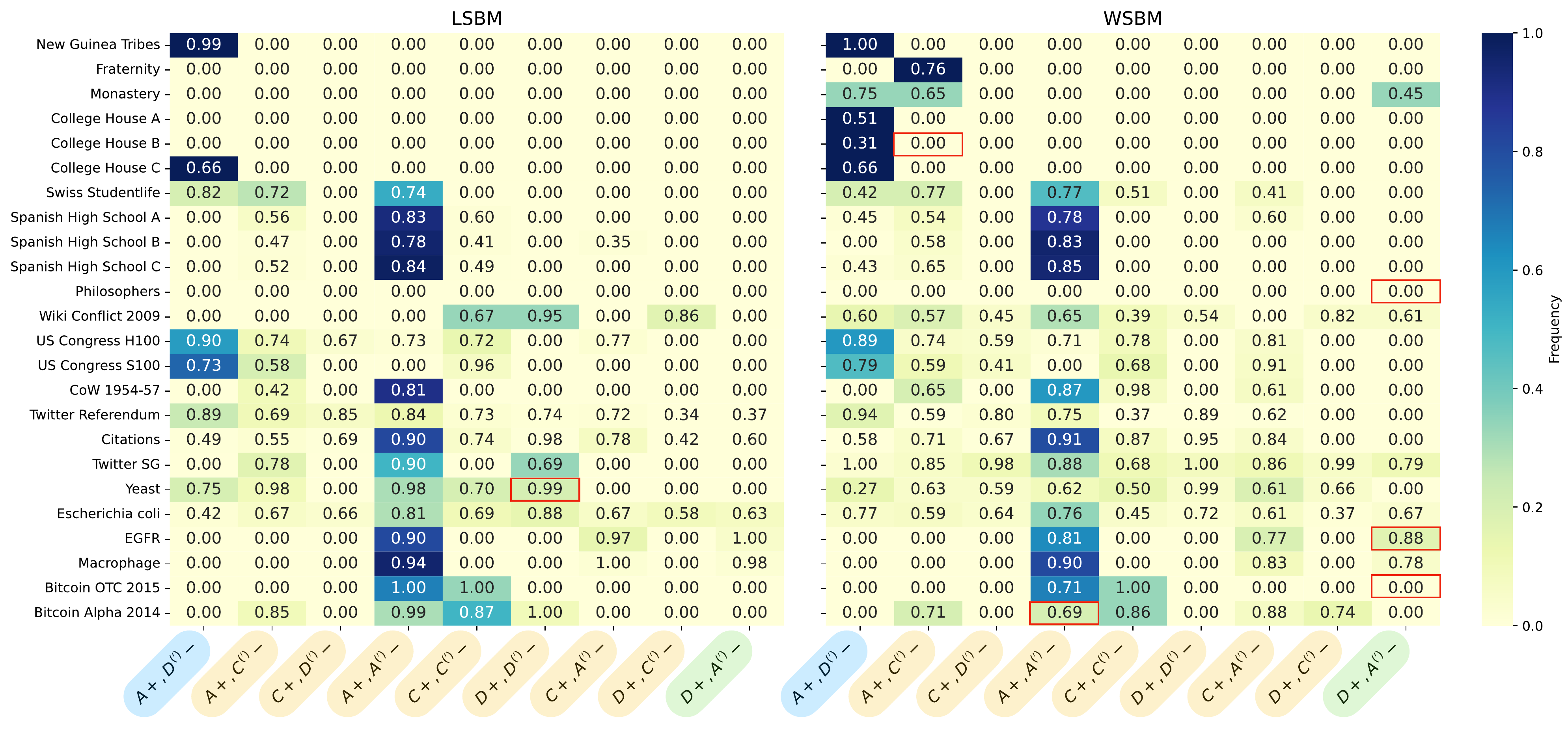}
\caption{\textbf{Details of the inferred structures for the survey by LSBM and WSBM.} 
The heatmaps illustrate the frequency and stability of nine fundamental interaction types. The background color intensity represents the interaction frequency (normalized from 0.0 to 1.0), while the numerical annotations denote the mean robustness of the type. 
We compared the dominant types identified in our survey with those from the posterior samplings (Table~\ref{tab:LSBM}-\ref{tab:WSBM}), and the inconsistent types were marked by red rectangles.
For networks with a dominant-type certainty exceeding 0.5, the consistency rate was 100\% for LSBM and 88\% for WSBM. When considering all networks, the consistency rates were 96\% for LSBM and 79\% for WSBM.
In cases of inconsistency, two main factors were identified. First, some discrepancies arise from low certainty in the posterior sampling (\textit{Yeast} for LSBM, \textit{College House B} and \textit{Philosophers} for WSBM), which suggests there are alternative community structures. Second, in other cases (\textit{ERFG}, \textit{Bitcoin OTC 2015} and \textit{Bitcoin Alpha 2014}), the survey-identified type corresponds to the second most frequent type in the posterior distribution, with a frequency very close to that of the dominant type.
}
\label{fig:S6}
\end{figure}

\clearpage

\begin{table}[]
\caption{\textbf{Details of the inferred structures for the Spanish High School by nested WSBM.} The table shows hierarchical community structure results for consensus partition and posterior sampling (Methods~\ref{ssec:sbm}). The consensus columns report the effective number of groups $B_e$, and the frequency of interaction types from the consensus partition. The sampling columns report the mean and standard deviation of the effective number of groups $\langle B_e \rangle$ and standard deviation, and the certainty of the interaction types. The dominant type of Level 1 is $(A^+, A^{(')-})$, which confirms the results in Fig.~\ref{fig:spanish}.}
\label{tab:spanish}
\resizebox{\textwidth}{!}{%
\begin{tabular}{lllllllllll}
\hline
\multicolumn{11}{c}{Consensus}                                  \\ \hline
Hierarchy & $B_e$ & $(A^+, D^{(')-})$ & $(A^+, C^{(')-})$ & $(C^+, D^{(')-})$ & $(A^+, A^{(')-})$ & $(C^+, C^{(')-})$ & $(D^+, D^{(')-})$ & $(C^+, A^{(')-})$ & $(D^+, C^{(')-})$ & $(D^+, A^{(')-})$ \\
Level 0   &    9.73   &        0.04           &          0.07         &     0.0              &        0.86           &       0.02            &         0.0          &          0.02         &           0.0        &         0.0          \\
Level 1   &   3.90    &        0.2           &        0.2           &       0.1            &        0.50           &         0.0          &       0.0            &         0.0          &        0.0           &          0.0         \\
Level 2   &    1.90   &        0.0           &       0.0            &       0.0            &          1.0         &        0.0           &        0.0           &           0.0        &          0.0         &         0.0          \\ \hline
\multicolumn{11}{c}{Sampling}                                   \\ \hline
Hierarchy & $\langle B_e \rangle$ & $(A^+, D^{(')-})$ & $(A^+, C^{(')-})$ & $(C^+, D^{(')-})$ & $(A^+, A^{(')-})$ & $(C^+, C^{(')-})$ & $(D^+, D^{(')-})$ & $(C^+, A^{(')-})$ & $(D^+, C^{(')-})$ & $(D^+, A^{(')-})$ \\
Level 0   &   9.92(36)      &       0.0          &         0.0          &            0.0       &          1.0         &        0.0           &        0.0           &          0.0         &        0.0           &        0.0           \\
Level 1   &   3.74(50)    &         0.0          &         0.05          &              0.0     &         0.94          &         0.0          &        0.0           &          0.0         &        0.0           &          0.0         \\
Level 2   &   1.95(61)    &         0.0          &         0.13          &           0.02        &           0.66        &          0.03         &      0.0             &         0.0          &         0.0          &       0.0  
\\ \hline
\end{tabular}%
}
\end{table}

\begin{table}[]
\centering
\caption{\textbf{Details of the inferred structures for the TwitterSG by nested WSBM.} The table shows hierarchical community structure results for consensus partition and posterior sampling (Methods~\ref{ssec:sbm}). 
The consensus columns report the effective number of groups $B_e$, and the frequency of interaction types from the consensus partition. The sampling columns report the mean and standard deviation of the effective number of groups $\langle B_e \rangle$ and standard deviation, and the certainty of the interaction types. 
We note that the posterior distribution of partitions exhibits a diffuse structure, suggesting multiple competing explanations for the network. To accurately capture this complexity, we identify and summarize the distinct modes of the distribution. Each mode represents a statistically significant, internally consistent clustering of the network’s organization.
The top levels of Modes 0 and 2 support balanced dominant types, while Modes 1 and 3 support unbalanced dominant types.
The results in Fig.~\ref{fig:cases}a show potential mesoscale structure of Modes 1 and 3.}
\label{tab:twitter}
\resizebox{\textwidth}{!}{%
\begin{tabular}{llllllllllll}
\hline
\multicolumn{12}{c}{Consensus} \\ \hline
Hierarchy & Mode & $B_e$ & $(A^+, D^{(')-})$ & $(A^+, C^{(')-})$ & $(C^+, D^{(')-})$ & $(A^+, A^{(')-})$ & $(C^+, C^{(')-})$ & $(D^+, D^{(')-})$ & $(C^+, A^{(')-})$ & $(D^+, C^{(')-})$ & $(D^+, A^{(')-})$ \\ \hline
Level 0 & 0 & 5.18 & 0.17 & 0.14 & 0.06 & 0.00 & 0.06 & 0.11 & 0.11 & 0.14 & 0.11 \\
        & 1 & 5.22 & 0.14 & 0.14 & 0.06 & 0.08 & 0.03 & 0.08 & 0.11 & 0.06 & 0.19 \\
        & 2 & 5.17 & 0.20 & 0.16 & 0.00 & 0.07 & 0.09 & 0.16 & 0.07 & 0.09 & 0.07 \\
        & 3 & 7.17 & 0.12 & 0.04 & 0.05 & 0.28 & 0.04 & 0.01 & 0.05 & 0.03 & 0.15 \\ \hline
Level 1 & 0 & 1.96 & 1.00 & 0.00 & 0.00 & 0.00 & 0.00 & 0.00 & 0.00 & 0.00 & 0.00 \\
        & 1 & 2.04 & 0.33 & 0.00 & 0.00 & 0.00 & 0.00 & 0.33 & 0.00 & 0.00 & 0.00 \\
        & 2 & 1.99 & 0.67 & 0.00 & 0.00 & 0.33 & 0.00 & 0.00 & 0.00 & 0.00 & 0.00 \\
        & 3 & 5.38 & 0.10 & 0.10 & 0.00 & 0.29 & 0.14 & 0.00 & 0.19 & 0.00 & 0.05 \\ \hline
Level 2 & 0 & 1.00 & 0.00 & 0.00 & 0.00 & 0.00 & 0.00 & 0.00 & 0.00 & 0.00 & 0.00 \\
        & 1 & 1.97 & 0.00 & 0.00 & 0.00 & 0.00 & 1.00 & 0.00 & 0.00 & 0.00 & 0.00 \\
        & 2 & 1.95 & 1.00 & 0.00 & 0.00 & 0.00 & 0.00 & 0.00 & 0.00 & 0.00 & 0.00 \\
        & 3 & 2.63 & 0.00 & 0.00 & 0.00 & 0.33 & 0.67 & 0.00 & 0.00 & 0.00 & 0.00 \\ \hline
\multicolumn{12}{c}{Sampling} \\ \hline
Hierarchy & Mode & $\langle B_e \rangle$ & $(A^+, D^{(')-})$ & $(A^+, C^{(')-})$ & $(C^+, D^{(')-})$ & $(A^+, A^{(')-})$ & $(C^+, C^{(')-})$ & $(D^+, D^{(')-})$ & $(C^+, A^{(')-})$ & $(D^+, C^{(')-})$ & $(D^+, A^{(')-})$ \\ \hline
Level 0 & 0 & 8.91(29) & 0.65 & 0.00 & 0.00 & 0.00 & 0.00 & 0.00 & 0.00 & 0.00 & 0.34 \\
        & 1 & 9.00(00) & 0.27 & 0.02 & 0.00 & 0.00 & 0.00 & 0.27 & 0.06 & 0.00 & 0.38 \\
        & 2 & 9.00(00) & 0.65 & 0.01 & 0.00 & 0.00 & 0.00 & 0.00 & 0.00 & 0.01 & 0.33 \\
        & 3 & 13.0(00) & 0.00 & 0.00 & 0.00 & 1.00 & 0.00 & 0.00 & 0.00 & 0.00 & 0.00 \\ \hline
Level 1 & 0 & 2.49(64) & 0.94 & 0.00 & 0.01 & 0.00 & 0.00 & 0.00 & 0.00 & 0.04 & 0.00 \\
        & 1 & 3.22(41) & 0.77 & 0.21 & 0.02 & 0.00 & 0.00 & 0.00 & 0.00 & 0.00 & 0.00 \\
        & 2 & 3.25(52) & 0.84 & 0.15 & 0.00 & 0.00 & 0.02 & 0.00 & 0.00 & 0.00 & 0.00 \\
        & 3 & 7.02(14) & 0.00 & 0.00 & 0.00 & 0.75 & 0.00 & 0.00 & 0.25 & 0.00 & 0.00 \\ \hline
Level 2 & 0 & 1.45(54) & 0.14 & 0.00 & 0.08 & 0.00 & 0.21 & 0.00 & 0.00 & 0.00 & 0.00 \\
        & 1 & 2.02(14) & 0.12 & 0.00 & 0.01 & 0.00 & 0.87 & 0.00 & 0.00 & 0.00 & 0.00 \\
        & 2 & 2.22(47) & 0.57 & 0.00 & 0.19 & 0.00 & 0.22 & 0.00 & 0.00 & 0.00 & 0.00 \\
        & 3 & 3.06(24) & 0.00 & 0.00 & 0.00 & 0.00 & 1.00 & 0.00 & 0.00 & 0.00 & 0.00 \\ \hline
\end{tabular}%
}
\end{table}

\begin{table}[]
\caption{\textbf{Details of the inferred structures for the USA Congress by nested LSBM.} The table shows hierarchical community structure results for consensus partition and posterior sampling (Methods~\ref{ssec:sbm}). The consensus columns report the frequency of interaction types from the consensus partition for each fixed number of groups $B$. The sampling columns report the certainty of the interaction types for each fixed number of groups $B$. The primary interaction type is balanced, incorporating core-periphery alongside other types, which confirms the findings in Fig.~\ref{fig:cases}b.}
\label{tab:congress}
\resizebox{\textwidth}{!}{%
\begin{tabular}{lllllllllll}
\hline
\multicolumn{10}{c}{Consensus}                                  \\ \hline
Fixed $B$ & $(A^+, D^{(')-})$ & $(A^+, C^{(')-})$ & $(C^+, D^{(')-})$ & $(A^+, A^{(')-})$ & $(C^+, C^{(')-})$ & $(D^+, D^{(')-})$ & $(C^+, A^{(')-})$ & $(D^+, C^{(')-})$ & $(D^+, A^{(')-})$ \\
2   &    0.71   &        0.11           &          0.04         &     0.01              &        0.10           &       0.0            &         0.03          &          0.01         &           0.0               \\
3   &   0.68    &        0.14           &        0.02           &       0.01            &        0.10           &         0.10          &       0.04            &         0.0          &        0.0          \\
4   &    0.68   &        0.13           &       0.04            &       0.0            &          0.13         &        0.0           &        0.03           &           0.0        &          0.0          \\ 
5   &    0.71   &        0.13           &       0.02            &       0.0            &          0.11         &        0.0           &        0.03           &           0.0        &          0.0          \\ 
\hline
\multicolumn{10}{c}{Sampling}                                  
\\ \hline
Fixed $B$ & $(A^+, D^{(')-})$ & $(A^+, C^{(')-})$ & $(C^+, D^{(')-})$ & $(A^+, A^{(')-})$ & $(C^+, C^{(')-})$ & $(D^+, D^{(')-})$ & $(C^+, A^{(')-})$ & $(D^+, C^{(')-})$ & $(D^+, A^{(')-})$ \\
2   &    0.70   &        0.12           &          0.03         &     0.01              &        0.10           &       0.0            &         0.04          &          0.0         &           0.0               \\
3   &   0.70    &        0.13           &        0.03           &       0.01            &        0.10           &         0.0          &       0.04            &         0.0          &        0.0          \\
4   &    0.71   &        0.12           &       0.04            &       0.0            &          0.11         &        0.0           &        0.03           &           0.0        &          0.0          \\ 
5   &    0.70   &        0.13           &       0.03            &       0.01            &          0.10         &        0.0           &        0.03           &           0.0        &          0.0          \\ 
\hline
\end{tabular}%
}
\end{table}

\begin{table}[]
\caption{\textbf{Details of the inferred structures for the CoW by nested WSBM.} The table shows hierarchical community structure results for consensus partition and posterior sampling (Methods~\ref{ssec:sbm}). The consensus columns report the effective number of groups $B_e$, and the frequency of interaction types from the consensus partition. The sampling columns report the mean and standard deviation of the effective number of groups $\langle B_e \rangle$ and standard deviation, and the certainty of the interaction types. The dominant type of Level 1 is $(A^+, C^{(')-})$, which confirms the results in Fig.~\ref{fig:cases}c.}
\label{tab:CoW}
\resizebox{\textwidth}{!}{%
\begin{tabular}{lllllllllll}
\hline
\multicolumn{11}{c}{Consensus}                                  
\\ \hline
Hierarchy & $B_e$ & $(A^+, D^{(')-})$ & $(A^+, C^{(')-})$ & $(C^+, D^{(')-})$ & $(A^+, A^{(')-})$ & $(C^+, C^{(')-})$ & $(D^+, D^{(')-})$ & $(C^+, A^{(')-})$ & $(D^+, C^{(')-})$ & $(D^+, A^{(')-})$ \\
Level 0   &    4.72   &        0.2           &          0.1         &     0.1              &        0.3           &       0.1            &         0.1          &          0.0         &           0.0        &         0.0          \\
Level 1   &   2.89    &        0.0           &        0.67          &       0.33            &        0.0           &         0.0          &       0.0            &         0.0          &        0.0           &          0.0         \\
\hline
\multicolumn{11}{c}{Sampling}                                   
\\ \hline
Hierarchy & $\langle B_e \rangle$ & $(A^+, D^{(')-})$ & $(A^+, C^{(')-})$ & $(C^+, D^{(')-})$ & $(A^+, A^{(')-})$ & $(C^+, C^{(')-})$ & $(D^+, D^{(')-})$ & $(C^+, A^{(')-})$ & $(D^+, C^{(')-})$ & $(D^+, A^{(')-})$ \\
Level 0   &   4.68(20)      &       0.19          &         0.03          &            0.0       &          0.78         &        0.0           &        0.0           &          0.0         &        0.0           &        0.0           \\
Level 1   &   2.81(35)    &         0.0          &         0.76          &              0.01     &         0.22          &         0.0          &        0.01           &          0.0         &        0.0           &          0.0  
\\ \hline
\end{tabular}%
}
\end{table}
\newpage
\clearpage

\section{Derivation of Eq.~(\ref{eq.sw})}
\label{supp:derivation}

To derive the description length of the uncorrelated edge signs, $S(W)$, we model the sign of each undirected edge as an independent, identically distributed random variable governed solely by the global probability of a positive edge, $p$. Given the observed network $W$, the conditional likelihood is
\begin{equation}
P(W|p) = \prod_{ij \in E} p^{\delta_{W_{ij},+1}} (1-p)^{\delta_{W_{ij},-1}} = p^{L^+} (1-p)^{L^-},
\end{equation}
where $L^+$ is the number of positive edges, $L^-$ is the number of negative edges, and $L = L^+ + L^-$ is the total number of undirected edges.

In the Minimum Description Length (MDL) framework, we avoid fitting a specific point estimate for $p$ and instead compute a marginal likelihood. Assuming a uniform, non-informative prior $P(p) = 1$ for $p \in [0, 1]$, we integrate over $p$ to obtain
\begin{equation}
P(W) = \int_{0}^{1} p^{L^+} (1-p)^{L^-} dp.
\end{equation}
This integral is exactly the Beta function $(\alpha, \beta)$,  with parameters $\alpha = L^+ + 1$ and $\beta = L^- + 1$. Expressing this via the Gamma function, $\Gamma(n) = (n-1)!$, yields
\begin{equation}
P(W) = \frac{\Gamma(L^+ + 1)\Gamma(L^- + 1)}{\Gamma(L + 2)} = \frac{L^+! L^-!}{(L + 1)!}.
\end{equation}
To simplify, we expand the denominator as $(L+1)! = (L+1)L!$ and recognize the inverse of the binomial coefficient, $\binom{L}{L^+}^{-1} = \frac{L^+! L^-!}{L!}$, leading to
\begin{equation}
P(W) = \frac{1}{(L+1) \binom{L}{L^+}}.
\end{equation}
Finally, Eq.~(\ref{eq.sw}) is recovered computing $S(W)=-\ln P(W)$.

\begin{figure}[ht!]
\centering
\includegraphics[width=1\linewidth]{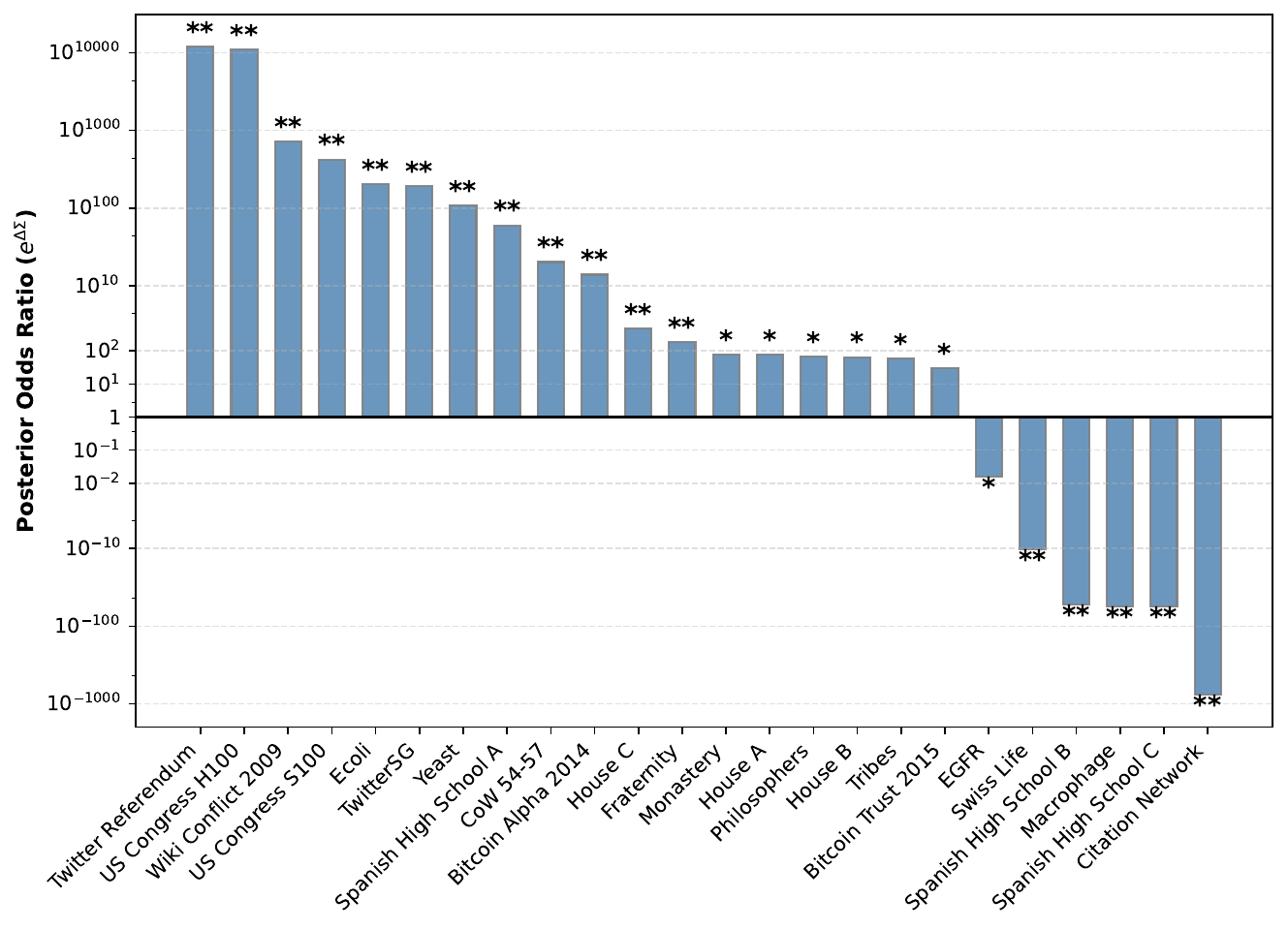}
\caption{\textbf{Structural information gained by incorporating edge signs across empirical networks.}
The y-axis displays the posterior odds ratio ($e^{\Delta \Sigma}$), quantifying how many times more likely the signed generative model is compared to the unsigned null hypothesis for the entire network (see Methods Sec.~\ref{model selection}). A symmetric log-scaled y-axis is utilized to ensure visual balance between positive evidence (ratios $> 1$, favoring the signed model) and negative evidence (ratios $< 1$, favoring the unsigned hypothesis), centered around a solid baseline at 1. Across the majority of datasets, edge signs act as a highly compressible generative signal. To highlight the statistical significance, datasets are marked with a single asterisk ($\ast$) if the posterior odds ratio favors either model by a factor exceeding 10 ($e^{|\Delta \Sigma|} > 10$), and a double asterisk ($\ast\ast$) if the factor exceeds 100 ($e^{|\Delta \Sigma|} > 100$).}
\label{fig:S7}
\end{figure}

\begin{figure}[ht!]
\centering
\includegraphics[width=1\linewidth]{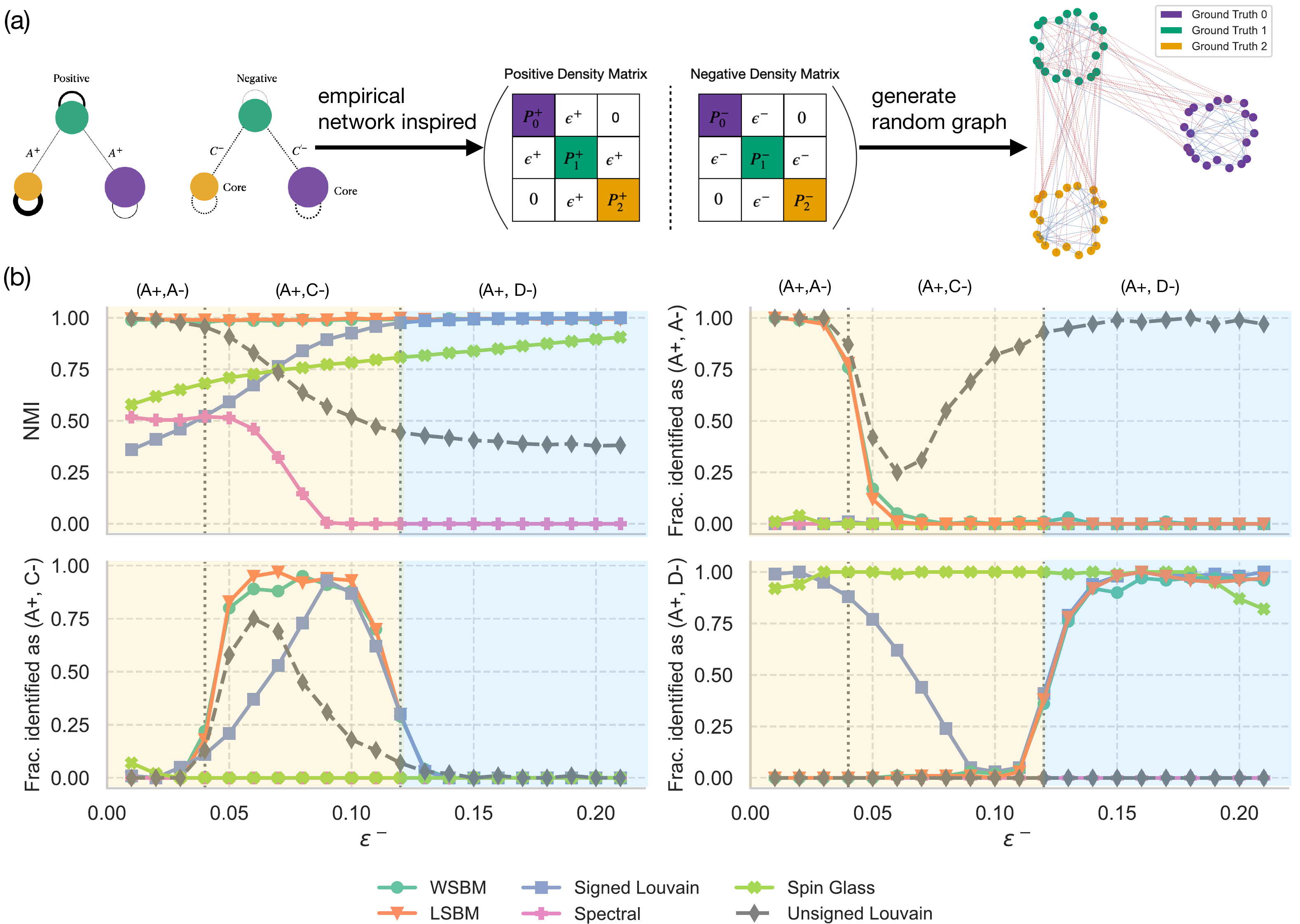}
\caption{\textbf{Heuristic methods lead to biased partitions in unbalanced synthetic networks.}
(a) We generate networks from a model consisting of three groups of equal size ($N/3=60$ nodes each), with block probability matrices inspired by the empirical density matrices of CoW studied in Fig.~\ref{fig:cases}: intra-group positive probabilities are asymmetric across blocks $P_i^+=(0.16, 0.08, 0.16)$, intra-group negative probabilities are non-negligible $P_i^-=(0.04, 0.12, 0.04)$, inter-group positive probability is fixed at $\epsilon^+ = 0.01$, and variable inter-group negative probability $\epsilon^-$. For $\epsilon^-<0.04$ the network is dominated by pure assortative $(A^+, A^{(')-})$ relationships, for $0.04 < \epsilon^- <0.12$ by the unbalanced $(A^+, C^{(')-})$ relationship, and for $\epsilon^->0.12$ by the balanced relationship $(A^+, D^{(')-})$.
(b) Performance of different methods as a function of $\varepsilon^-$. The four plots show the normalized mutual information (NMI) between detected and ground-truth partitions and fraction of networks in which each method identifies the dominant interaction type as $(A^+, A^{(')-})$, $(A^+, C^{(')-})$, and $(A^+, D^{(')-})$. Results are averaged across 100 independent random networks. The two structural transition are marked by vertical dotted lines, at $\epsilon^- = 0.04$ and $\epsilon^- = 0.12$.
Across all regimes, both SBMs correctly track the ground truth, maintaining high NMI throughout. Signed Louvain performs well only in the $(A^+, D^{(')-})$ regime. (Unsigned) Louvain performs well only in the $(A^+, A^{(')-})$ regime, where ignoring edge signs does not obscure the community structure. Both Louvain and Signed Louvain do not work well in the $(A^+, C^{(')-})$ regime. 
}
\label{fig:S8}
\end{figure}

\newpage
\clearpage

\end{document}